\documentclass[a4paper,11pt]{article}
\pdfoutput=1  

\usepackage{jcappub} 

\usepackage[T1]{fontenc}  
\usepackage{bm}
 
\title{\boldmath Beyond the Poles in Attractor Models of Inflation}

 \author{Sotirios Karamitsos}
 \affiliation{Consortium for Fundamental Physics, Physics Department,
Lancaster University\\ Lancaster LA1 4YB, United Kingdom}
 
\emailAdd{s.karamitsos@lancaster.ac.uk}

 \abstract{
 We offer a geometric interpretation of attractor theories with singular kinetic terms as a union of multiple canonical models. We demonstrate that different domains (separated by poles) can drastically differ in their phenomenology. We illustrate this with the help of a ``master model'' that leads to distinct predictions depending on which side of the pole the field evolves before examining the more realistic example of $\alpha$-attractor models. Such models lead to quintessential inflation within the poles when featuring an exponential potential. However, beyond the poles, we discover a novel behaviour: the scalar field responsible for the early-time acceleration of the Universe may reach the boundary of the field-space manifold, indicating that the theory is incomplete and that a boundary condition must be imposed in order to determine its late-time behaviour. If the evolution of the field is arrested before this happens, however, we discover that quintessence can be achieved without a potential offset. Turning to multifield models with singular kinetic terms, we see that poles generalise straightforwardly to singular curves, which act as ``model walls'' between distinct pole-free inflationary models. As an example, we study a simple two-field $\alpha$-attractor-inspired model, whose evolution of isocurvature perturbations is sensitive to where the non-canonical field begins its trajectory. We finally discuss initial conditions in attractor theories, where the existence of multiple disconnected canonical models implies that we must make a fundamental choice: in which domain we impose a distribution for the inflaton in order to then determine the likelihood of inflation.
}
 
\keywords{inflation, physics of the early universe, modified gravity, particle physics -- cosmology connection}

\begin{document}
\maketitle
\flushbottom
\section{Introduction}
\label{intro}

There is no questioning that inflation \cite{Guth:1980zm,Linde:1981mu,Hawking:1982cz} is one of the best explanations for the standard problems of standard Big Bang cosmology. One of these is the well-known \emph{flatness} problem, which stems from the observation that the early curvature of the Universe must have been fine-tuned to the order of $\sim 10^{-55}$ to match current observations \cite{Aghanim:2018eyx}. Another is the \emph{horizon} problem, which asks why the temperature profile of the cosmic microwave background (CMB) is nearly uniform to within a few $\mu {\rm K}$ while the horizon at the time of last scattering corresponds to about $1^\circ$ in the sky today. 

A period of accelerated expansion in the early Universe resolves both the flatness and horizon problems. However, one of the most celebrated predictions of inflation is the large scale structure of the Universe today \cite{Ratra:1987rm,Mukhanov:1990me,Lyth:1998xn}, which is \emph{nearly} scale-invariant, adiabatic, and Gaussian \cite{Akrami:2018odb}. These predictions are \emph{generic}; a sufficiently prolonged period of early-time acceleration is going to both resolve the classic cosmological puzzles while simultaneously generating primordial perturbations that become the seeds for large scale structure. Unfortunately, the generic nature of these predictions is a double-edged sword for the theory of inflation, making it a significant challenge to discriminate between potential models. 

A particularly popular paradigm in inflation is found in \emph{scalar-tensor} models, which feature a bosonic field~$\phi$ coupled to gravity with an arbitrary potential. The most general scalar-tensor Lagrangian can be specified by three model functions: a \emph{non-minimal} coupling~$f(\phi)$ between the scalar field and the Ricci scalar, a multiplicative \emph{non-canonical} factor~$k(\phi)$ for the kinetic term,\footnote{In the literature, ``non-canonical'' usually refers to the presence of higher-derivative kinetic terms. In this paper, we restrict ourselves to theories quadratic in derivatives, and so we shall refer to the so-called ``designer'' kinetic terms $k(\phi) (\partial\phi)^2/2$ as non-canonical when $k(\phi) \ne 1$.} and finally, the potential $V(\phi)$. The interaction of the field with the matter part of the Lagrangian is ignored; while it may be important during reheating, the energy density of the Universe is dominated by the scalar field during inflation.

Scalar-tensor models offer ample opportunities for model-bulding: indeed, if we carefully select the model functions, it is possible to generate an almost unlimited array of models. Well-motivated examples of such models include power-law inflation~\cite{Lucchin:1984yf,Halliwell:1986ja}, natural inflation~\cite{Freese:1990rb}, axion inflation~\cite{Linde:1991km,Pajer:2013fsa}, hybrid inflation~\cite{Linde:1993cn}, false vacuum inflation \cite{Copeland:1994vg}, 
brane inflation~\cite{Burgess:2001fx}, hilltop inflation~\cite{Boubekeur:2005zm}, and supergravity inflation~\cite{Yamaguchi:2011kg}. 

Recently, there has been much interest in a wide class of non-canonical inflationary models known as \emph{attractor models}. An early instance of such models were the conformal attractors~\cite{Kallosh:2013hoa}, later extended to the now well-known $\alpha$-\emph{attractors}~\cite{Kallosh:2013wya,Kallosh:2013xya}. These models make predictions that are insensitive to the details of the inflationary potential $V(\phi)$ in the limit of small $\alpha$~\cite{ Kallosh:2013yoa,Kallosh:2014rga,Kallosh:2015lwa, Carrasco:2015pla,Carrasco:2015rva,Scalisi:2015qga,Rinaldi:2015yoa,Roest:2015qya,
Linde:2015uga,Kallosh:2016gqp} (justifying the ``attractor'' moniker), as well as remaining robust under renormalisation group flow corrections \cite{Fumagalli:2016sof}. Models where $f(\phi) \propto 1 + \xi \sqrt{V(\phi)}$ (such as Higgs inflation) or $f(\phi) \propto \xi \sqrt{V(\phi)}$ (induced gravity inflation) also feature attractor behaviour, and are known as $\xi$-\emph{attractors}. In the limit $\xi\gg1$, their predictions are also insensitive to the form of~$V(\phi)$~\cite{Kallosh:2013tua}.

The unifying feature of attractor models is that their predictivity is largely determined by the features of their kinetic term~\cite{Galante:2014ifa}. If the kinetic term features a pole, redefining the field $\phi$ via $\varphi(\phi)$ such that the kinetic term becomes canonical results in a ``stretching'' of the potential. No matter what the original potential is, the canonical potential $U(\varphi) = V(\varphi(\phi))$ acquires a shift symmetry at large field values, making it easy to see how the attractor behaviour is realised. In general, the observables are found to depend strongly on the \emph{order} and \emph{residue} of the kinetic pole. This paradigm is referred to as \emph{pole inflation}~\cite{Broy:2015qna,Terada:2016nqg,Szydlowski:2017evb, Dias:2018pgj}. The presence of poles in the kinetic term can be physically motivated in various ways: these include superconformal symmetry breaking on a K\"ahler manifold with curvature $1/\alpha$ \cite{Kallosh:2013wya,Kallosh:2013xya,Carrasco:2015uma} and alternative theories of gravity \cite{Kim:2016bem, Odintsov:2018qyy,Odintsov:2016vzz}.

While pole inflation is a useful formalism for studying theories with poles, it is important to remember that the presence of non-canonical kinetic terms does not introduce any new physics, much like introducing a non-minimal coupling does not lead to a distinct class of models, at least at the tree level~\cite{ Gasperini:1993hu,faraoni98,Flanagan:2004bz,Chiba:2013mha,Steinwachs:2013tr,
Kamenshchik:2014waa,Postma:2014vaa,Jarv:2014hma,Domenech:2015qoa,Burns:2016ric,Karam:2017zno,Karam:2018squ}. Indeed, we are free to conformally scale the metric and reparametrise the inflaton without changing the physical content of a single-field models. Therefore, referring to a model as (non)minimal or (non)canonical is misleading; it is the \emph{representation} of a model (i.e. its Lagrangian) that can be (non)minimal or (non)canonical. In fact, single-field models can be fully specified with just one functional parameter $U(\varphi)$~\cite{Jarv:2016sow}, while multifield models can be specified with the help of $U(\varphi)$ and the multifield kinetic mixing term $G_{AB}$~\cite{Karamitsos:2017elm}. 

The one-to-one correspondence between models related by a frame transformation is broken in the presence of the aforementioned poles in the kinetic term. For instance, when canonicalising the $\alpha$-attractor Lagrangian, we find that the domain of the canonical field expressed in terms of the non-canonical field is not the entire real line. This indicates that when poles are present in the Lagrangian, more than a single definition of the canonical field is required in order to reproduce the original non-canonical action. Attractor models can thus be viewed as a collection of distinct single-field models. As such, the phenomenology of scalar-tensor theories with singular kinetic terms will in general depend on the interval in which the non-canonical field is found. This raises important questions for the choice of initial conditions; since poles cannot be crossed, choosing a domain for the non-canonical field amounts to selecting a manifestly different canonical potential. Examining the domain beyond the poles, which is usually ignored in the literature, is therefore a potential avenue for novel model building.
 
 This paper is laid out as follows: in Section~\ref{sttheories}, we outline how scalar-tensor theories behave under frame transformations, examining their manifold structure and showing how they can viewed as unions of pole-free theories. In Section~\ref{singlefieldpoles}, we provide an overview of single-field pole inflation and its attractor behaviour. We proceed to examine how the phenomenology of attractor models may depend on the domain in which the field evolves. We illustrate the role of canonical Lagrangians as the building blocks of single-field attractor models by constructing a kinetically singular model  which features two completely distinct models on different sides of the pole. We move on to examine $\alpha$-attractors Lagrangian in Section~\ref{alphaattractors}, focusing on models with an exponential potential. Such a setup corresponds to quintessential inflation between the poles. However, beyond the poles, we find that the inflaton field may reach the end of the field space manifold in finite time while still accounting for the generation of primordial perturbations. If this does not occur, we find instead that quintessence may also be achieved beyond the poles. We return to multifield models in Section~\ref{modelwalls}. With multiple scalar fields, poles generalise to \emph{singular curves} where the kinetic terms become singular. We examine the conditions under which these curves cannot be crossed, and demonstrate that they act as ``model walls'' in a way analogous to poles for single-field inflation. This means that they divide the field-space manifold into different regions, each of which corresponds to a distinct singularity-free model. We illustrate the difference in phenomenology beyond the poles in the context of entropy perturbations. In particular, we consider a two-field $\alpha$-attractor model in which entropy generation can be generically amplified outside of the hyperbolic domain. Finally, we conclude and discuss our findings in Section~\ref{conclusions}.

\section{Scalar-tensor theories as manifolds}
\label{sttheories}

In this section, we will specify the Lagrangian for single-field scalar-tensor theories. This is going to lay the foundation for the construction of the \emph{Lagrangian space}, whose points represent different representations of scalar-tensor theories. We demonstrate how conformal transformations and field reparametrisations can be used to recast the same model in different frames, giving rise to \emph{orbits} whose points represent the same underlying theory. Using the language of differential geometry, we will motivate the notion of a field space and demonstrate that when a Lagrangian features poles, more than one canonical field is required to fully parametrise the corresponding theory.

We begin by writing the general single-field scalar-tensor Lagrangian as follows:
 \begin{align}
\label{actionEFnoncan}
(-g)^{-1/2} \mathcal{L} =  - \frac{f({  \phi})}{2}R + \frac{k ({ \phi})}{2} (\partial_\mu \phi )(\partial^\mu \phi ) - V_J({ \phi})  .
\end{align}
The model described by the Lagrangian in \eqref{actionEFnoncan} is said to be expressed in the \emph{Jordan frame}, which means that the standard Einstein gravity term $M_P^2 R$ has been promoted to $f({ \phi}) R$. The Lagrangian is specified by the following model functions: $f({ \phi}) $ is known as the non-minimal coupling,~$k ({ \phi})$ is the non-canonical coupling, and~$V_J({ \phi})$ is the Jordan frame inflationary potential. Furthermore, the Ricci scalar is denoted by $R$, Greek indices are contracted with the spacetime metric~$g_{\mu\nu}$, and $g= \det g_{\mu\nu}$. 

At first glance, it appears that the scalar-tensor ``model space'' has three functional degrees of freedom, since~$f({ \phi})$,~$k ({ \phi})$, and $V_J({ \phi})$ fully specifies the model. However, there are two transformations that allow us to express the model in a simpler way. We may eliminate the explicit appearance of $f({ \phi})$ via a \emph{conformal transformation}, which, as standard in cosmology, is a field-dependent rescaling of the metric:
\begin{align}
\label{conftrans}
g_{\mu\nu } \to \Omega({ \phi})^2 g_{\mu\nu}.
\end{align}
If we choose $\Omega^2 =M_P^2/f(\varphi)$, then the Lagrangian becomes~\cite{Flanagan:2004bz,Jarv:2014hma}
\begin{align}
\label{s}
(-g)^{-1/2} \mathcal{L} =  -\frac{M_P^2 R}{2} + \frac{G(\phi) }{2}(\partial_\mu \phi)(\partial^\mu \phi) -V_E(\phi),
\end{align}
where the new model functions are given by
\begin{align}
\label{fieldspacemetric}
G (\phi) &\equiv\   M_P^2 \left( \frac{ k(\phi) }{f(\phi) }   +\frac{3}{2}   \frac{ f'(\phi)^2 }{f(\phi)^2} \right),
\qquad
V_E (\phi) \equiv \frac{V_J(\phi)}{f(\phi)^2/M_P^4}.
\end{align}
In order to avoid tachyonic modes, we assume that $G(\varphi)>0$. The action is now minimally coupled; the interaction between the curvature and the field has been ``hidden'' in the kinetic term.

It is possible to further simplify the Lagrangian by reparametrising the fields as $\varphi  = \varphi ({ \phi})$ in a way that renders the kinetic term canonical:
\begin{align}
\label{multifieldredef}
\left(\frac{\partial\varphi }{\partial\phi }\right)^2 = G(\phi).
\end{align}
We will refer to the combined effect of a conformal transformation  and a field reparametrisation as a \emph{frame transformation}. Applying this procedure allows us to write the Lagrangian as follows:
\begin{align}
\label{actionEFcanmultifield}
(-g)^{-1/2} \mathcal{L} = -\frac{M_P^2 R}{2} +\frac{1}{2}(\partial_\mu \varphi )(\partial^\mu \varphi ) - U({  \varphi}) .
\end{align}
Here, we have inverted \eqref{multifieldredef} to write $\phi  = \phi  ({  \varphi})$, and
\begin{align}
U({  \varphi}) \equiv V_E({  \phi}({  \varphi})).
\end{align}
When $f(\varphi) = M_P^2$, such as in~\eqref{actionEFnoncan} and~\eqref{actionEFcanmultifield}, the theory is said to be in the \emph{Einstein frame}, although some authors reserve this term only for the latter case when the field is also canonical. In order to be unambiguous, we will say that the Lagrangian \eqref{actionEFnoncan} is written in the \emph{non-canonical} Einstein frame and the Lagrangian \eqref{actionEFcanmultifield} is written in the \emph{canonical} Einstein frame. We will refer to $U(\varphi)$ as the (Einstein frame) \emph{canonical potential}, as opposed to~$V_E(\phi)$, which we call the (Einstein frame) non-canonical potential, usually dropping the index and simply writing $V(\phi)$.

\begin{figure} 
  \centering
  \hspace{-1em}
\includegraphics[scale=1.2]{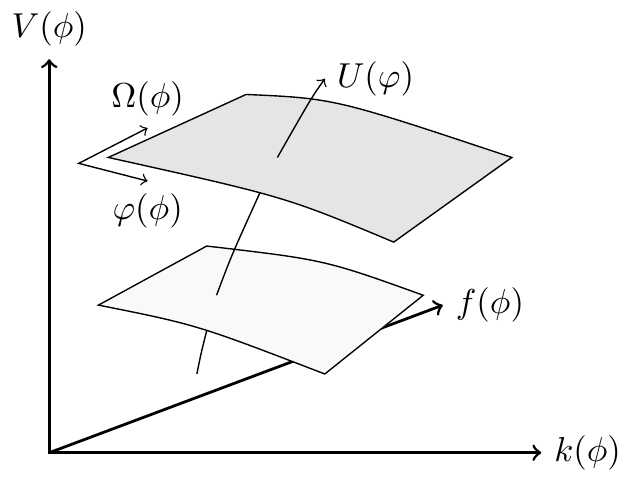}
  \caption{Equivalence classes of single-field scalar-tensor Lagrangians. Lagrangians on the same equivalence classes correspond to the same physical models (shown as distinct sheets).  }
  \label{fig:orbits}
\end{figure}

Applying a frame transformation does not leave the action invariant, which leads to the natural question of whether frame transformations are physically meaningful. This is known as the \emph{frame problem}, a long-standing debate whose origin can be traced to alternative theories of gravity \cite{Gasperini:1993hu,faraoni98}. It is now commonly accepted that Lagrangians related by frame transformations are different expressions of the same model, insofar as they produce the same observables, at least at the tree level \cite{ Flanagan:2004bz,Chiba:2013mha,Steinwachs:2013tr,
Kamenshchik:2014waa,Postma:2014vaa,Domenech:2015qoa,Burns:2016ric,Jarv:2014hma,Jarv:2016sow,
Karamitsos:2017elm,  Karam:2017zno,Karam:2018squ}.  Therefore, even if ostensibly it takes three functions $f(\phi), k(\phi)$ and $V(\phi)$  to specify a \emph{Lagrangian}, a physically distinct single-field \emph{model} can be fully specified by the canonical potential $U(\varphi)$. More rigorously, single-field scalar-tensor Lagrangians linked by a frame transformation belong to the same \emph{equivalence class}, corresponding to the same physical model.
This is shown schematically in Figure~\ref{fig:orbits}, which depicts the three-dimensional ``Lagrangian space'' whose members are the different Lagrangians $\mathcal{L}[f(\phi), k(\phi), V(\phi)]$. We can therefore see that the \emph{model space}\footnote{The model space in which physically distinct models live is none other than the \emph{quotient space} of the Lagrangian space defined with respect to the frame equivalence relation.} (in which physically distinct models live, as opposed to Lagrangians) has a lower dimension than the Lagrangian space. 
Indeed, this model space is a collection of all sheets $M_{U(\varphi)}$, specified just by the canonical potential $U(\varphi)$ and given by
\begin{align}
M_{U(\varphi)} = \left\{\mathcal{L} 
\left[  
\Omega ^{-2} M_P^2, \
\frac{\Omega ^{-2}}{K } 
\left[  1- 6M_P^2 \left(\frac{d \ln \Omega }{d\varphi}\right)^2   \right], \
\Omega^{-4} \, U(\varphi) 
 \right] \, \bigg| \,  \forall \Omega(\phi), K(\phi)
\right\} 
\end{align}
where $\Omega(\phi)>0$ and $K(\varphi)$ is invertible everywhere (and $K(\phi) = G(\phi) = \varphi'(\phi)^2$ when canonicalising the field).

The above discussion might give the impression that non-minimal and non-canonical models are in a sense trivial, since they can be fully classified with just a single function. However, as we noted just above, a very important condition that must be satisfied
 for a single canonical potential to fully capture the phenomenology of a general scalar-tensor Lagrangian is that the solution to the canonicalisation equation \eqref{multifieldredef} is one-to-one. 
  If the solution to \eqref{multifieldredef} is not one-to-one, it is not possible to invert~$\varphi(\phi)$, and defining $U(\varphi) = V(\phi(\varphi))$ is also not possible. ``Inverting'' a function that is not one-to-one results in a one-to-many mapping, which we may write as~$\phi_i (\varphi)$. This allows us to define a \emph{collection} of canonical potentials $U_i (\varphi) = U(\phi_i(\varphi))$, where the index $i$ specifies different intervals~$I_i$ on which the function~$\varphi(\phi)$ is one-to-one.
 
The formal solution to \eqref{multifieldredef} is 
\begin{align} 
\label{formsol}
  \varphi(\phi)= \int_{\phi_0}^\phi  d\phi' \sqrt{G(\phi')},
\end{align}
where $\phi_0$ is the arbitrary point which we choose to correspond to $\varphi = 0$ (usually taken to be~$\phi_0 = 0$). It is not difficult to see that this equation is one-to-one. However, if $G(\phi)$ is singular at some point, i.e. it features a \emph{pole}, it is not possible to integrate over it. If the pole is located at $\phi = \phi_p$, we find that $\varphi$ will run over the entire real line as $\phi \in I_- = (-\infty, \phi_p)$, as long as the arbitrary origin point satisfies $\phi_0<\phi_p$. Similarly,  $\varphi$ will run over the entire real line as $\phi \in I_+ = (\phi_p,\infty)$. As such, we find that a single canonical field (and hence potential) is not enough to fully describe a model whose kinetic term is singular.

In order to understand the relation between the canonical and non-canonical fields, we turn to the language of differential geometry. Much like a Lagrangian is simply a representation of an underlying model, $\varphi$ and $\phi$ are two distinct coordinate \emph{charts} parametrising a particular manifold which is known as the \emph{field space}. In order to see this behaviour more clearly, we turn our attention to multifield models. The general multifield Lagrangian can be written as follows in the Einstein frame:
\begin{align}
\label{minnoncan}
(-g)^{-1/2} \mathcal{L} =  - \frac{M_P^2}{2}R + \frac{G_{AB} ({ \bm \phi})}{2} (\partial_\mu \phi^A )(\partial^\mu \phi^B) - V({ \bm \phi}),
\end{align}
where $1\le A,B \le n$, field indices are summed over, and the fields $\phi^A$ are collectively denoted by ${\bm \phi}$. The relation between this Lagrangian and a manifold becomes manifest when looking at the equations of motion for a FLRW metric and homogeneous fields $\phi^A = \phi^A(t)$. These are
\begin{align}
\label{eom1}
\begin{aligned}
\ddot \phi^A &+ \Gamma^A_{BC} \dot\phi^B \dot\phi^C + 3 H \dot \phi^A + G^{AB} V_{,B} = 0,
\\
3 H^2 &= \frac{G_{AB}\dot\phi^A \dot\phi^B}{2} + V,
\end{aligned}
\end{align}
where the inverse matrix $G^{AB}$ is defined via $G^{AB}G_{BC} \equiv \delta^A_C$. 
The Hubble parameter is defined as usual through $H= \dot a/a$, the overdot denotes differentiation with $t$, $V_{,A} \equiv \partial V/\partial \phi^A$, and
\begin{align}
\Gamma^A_{BC} = \frac{G^{AD}}{2} \left(G_{DB,C} + G_{CD,B} -G_{BC,D}\right).
\end{align}
The first equation of \eqref{eom1} is strikingly similar to a geodesic equation with an additional drag term and a conservative force. We can therefore see that $G_{AB}$ takes on the role of the metric on this manifold, and we refer to it as the \emph{field-space metric}. 

Formally, a chart is a homeomorphism (continuous function with continuous inverse) from an open subset of a manifold to an open subset of Euclidean space. We can represent a chart  as the ordered pair $(\mathcal{L},\phi)$, where we slightly abuse notation to let $ \mathcal{L}$ also stand for the field-space manifold corresponding to the model described by the Lagrangian. What we have found is that a single canonical chart $(\mathcal{L},\varphi)$ is not enough to fully parametrise the entirety of the field space. In general, when the Lagrangian features poles, one canonical chart does not suffice to fully parametrise $\mathcal{L}$, even if the field-space manifold is one-dimensional. What we need is a collection of charts, known as an \emph{atlas}. In the example of a single-field Lagrangian featuring one pole, we may write
\begin{align}
\mathcal{L} = \mathcal{L}_- \cup \mathcal{L}_+.
\end{align}
We have decomposed the field space into two submanifolds (which are still field spaces), one to the left and one to the right of the pole. None of these field spaces feature any singularities and so may be parametrised with a single chart. Finally, we can parametrise $\mathcal{L}$ with a canonical atlas $\{( \mathcal{L}_-, \varphi_-),( \mathcal{L}_+, \varphi_+) \}$ that features only canonical charts.\footnote{For a canonical atlas, there is no overlap between the submanifolds $\mathcal{L}_-$ and $\mathcal{L}_+$, and therefore there is no need to define a transition map.} It is not difficult to extend this construction to models with an arbitrary number of singularities: the existence of $n$ poles means that $n+1$ canonical parametrisations are required.

In this section, we have shown how single-field canonical models can be thought of as the fundamental ``building blocks'' of models specified by non-canonical Lagrangians that feature poles. In order to make our treatment more concrete, we will turn our attention to single-field pole inflation, focusing on the differences in phenomenology that arise depending on which submanifold we chose to restrict ourselves to.

\section{Single-field pole inflation}
\label{singlefieldpoles}
 
We have seen that when the kinetic term of a single-field Lagrangian is singular, a canonical model is no longer sufficient for representing the entirety of the underlying theory.  Poles may arise in various ways: common examples include the hyperbolic geometry of the manifold on which inflation occurs~\cite{Kallosh:2013wya,Kallosh:2013xya,Carrasco:2015uma} and the modification of the gravity sector, such as via non-minimal couplings~\cite{Kim:2016bem, Odintsov:2018qyy} or $F(R)$ gravity \cite{Odintsov:2016vzz}.
No matter their origin, poles introduce subtleties in the picture painted in Figure~\ref{fig:orbits} and in the previous section.
In this section, we will review pole inflation \cite{Galante:2014ifa,Broy:2015qna,Terada:2016nqg} and its relation to single field canonical Lagrangians. We will explicitly demonstrate that in the presence of singularities in the kinetic term, canonicalising the field is a procedure that does not preserve the structure of the original Lagrangian beyond the poles. As such, since multiple canonical potentials are required to recover the original theory, there will be stark phenomenological differences depending on which submanifold the field evolves, at least away from the poles. We further illustrate this point by constructing a ``master Lagrangian'' with an order~2 pole that hides an arbitrary number of completely distinct theories. 

 \subsection{Overview of pole inflation}

We begin by writing down a Lagrangian whose kinetic term features a pole of order $p$. There may be other poles, but we focus on the pole of highest order. If the kinetic term has a single pole, then it is possible to write the Lagrangian close to the pole as
\begin{align}
\label{poleaction}
(-g)^{-1/2}\mathcal{L}=  - \frac{  R}{2} + \frac{\alpha_p}{2} \frac{(\partial_\mu \phi) (\partial^\mu\phi)}{|\phi  |^p}  - V(0) - V'(0) \phi,
\end{align}
where from now on we set $M_P = 1$ for ease of notation and we assume that the potential has a negative gradient (such that the field is pushed away from the pole when $\phi>0$). We have also shifted the field such that the pole coincides with $\phi = 0$.

It is not difficult to extract inflationary predictions from this model without canonicalising the potential. It suffices to use the extended expressions for the slow-roll parameters when not in the Einstein frame \cite{Burns:2016ric,Jarv:2016sow,Karamitsos:2017elm}:
\begin{align}
\label{srp}
\epsilon &= \frac{1}{2k(\phi)} \frac{V'(\phi)^2}{V(\phi)^2} ,
&
\eta &= \frac{\epsilon'(\phi)}{\epsilon(\phi)} \frac{1}{k(\phi)} \frac{V''(\phi)}{V(\phi)},
\end{align}
and the number of $e$-foldings is given by
\begin{align}
\label{efolds}
N (\phi) =  \int_{\phi_{\rm end}}^{\phi }d\phi' \  k(\phi')\frac{V(\phi')}{V'(\phi')}.
\end{align}
The value of the field at the end of inflation is given by $\epsilon(\phi_{\rm end}) = 1$.

Since the field will never actually reach $\phi = \phi_0$, we must be careful to distinguish whether we take the limit of $\phi\to \phi_0$ from the left or from the right depending on which side of the pole we wish to inflate on. Using the standard expressions for the scalar tilt and tensor-to-scalar ratio $n_\mathcal{R} = 1 -2 \epsilon+\eta$ and $r = 16\epsilon$, we find that the observables for the theory given in \eqref{poleaction} are
\begin{align}
\begin{aligned}
n_\mathcal{R} &= \frac{\alpha_p  \phi +v_0^2 \phi  \left(\alpha_p  \phi ^2+(p-3) \phi ^p\right)-v_0 \left(2 \alpha_p  \phi ^2+p \phi ^p\right)}{\alpha_p  \phi  (1-v_0 \phi )^2},
\\
r &= \frac{8 v_0^2 \phi ^p}{\alpha_p  \left(1 - v_0 \phi   \right)^2},
\end{aligned}
\end{align}
where $v_0 = \lim_{\phi \to 0} v(\phi)$ and $v(\phi)$ is defined as
\begin{align}
\label{defv}
v(\phi) =  -\frac{V'(\phi)  }{V(\phi)} = - \sqrt{2\epsilon} .
\end{align}
In general, in order to have a graceful exit from inflation, we require that $v_0>0$ in order for the field to roll away from the pole when $\phi>0$. Moreover, 
we note that most ``well-behaved'' potentials satisfy 
\begin{align}
\label{limnoneq}
 \lim_{\phi \to \phi_0^+ } \left\vert \frac{v(\phi)}{v(-\phi)}\right\vert  =  1.
\end{align}
Such potentials are continuously differentiable if they  also happen to be continuous. If they are not continuously differentiable, then it is not possible to set $v_0 = 1$ independent of which side we expand by rescaling $\phi$. As we shall demonstrate in the next subsection, this has stark implications for the phenomenology of the different domains of the theory (which are delimited by its poles).
 
Substituting the explicit form for $k(\phi)$ and $V(\phi)$ from \eqref{poleaction} into \eqref{srp} and \eqref{efolds}, we find that close to the pole, the number of $e$-foldings is given by
\begin{align}
N =  
\begin{cases}
\frac{\alpha_p }{  v_0 (p-1)} \left(\phi^{1-p}-\phi_{\rm end}^{1-p}\right) & (p\ne1),
\\
 -\frac{\alpha_p}{ v_0}\ln \frac{\phi }{\phi_{\rm end}} &  (p=1).
\end{cases}
\end{align}
We note that $\epsilon = 1$ can be used to find $\phi_{\rm end} = |2\alpha_p|^{1/p}$.

Substituting for the number of $e$-foldings, we find that the observables to lowest order are
 \begin{align}
\label{observables}
 n_\mathcal{R} &= 1 - \frac{ p}{(p-1) N}, & r &= \frac{8 v_0^2 }{\alpha_p } \left[ \frac{\alpha_p }{(p-1) v_0 N} \right]^{p/(p-1)}.
\end{align}
These expressions agree with those found in~\cite{Terada:2016nqg}, and they hold for $p\ne 1$. For $p=1$, the observables obtain a different form. For large $N$, we find:
 \begin{align}
\label{observables2}
\begin{aligned}
 n_\mathcal{R} &= 1  -\frac{v_0}{\alpha_p}  -8 v_0^2 e^{-\frac{ v_0}{\alpha_p}N},
 \\
 r &= 16 v_0^2 e^{-\frac{ v_0}{\alpha_p} N}.
\end{aligned}
\end{align}
We therefore recover the well-known results from pole inflation: the spectral tilt depends solely on the order of the pole when $p\ne 1$, whereas the tensor-to-scalar ratio depends on the residue and the order (and the potential, unless $p=2$).
 
\begin{figure} 
  \centering
  \hspace{-1em}
\includegraphics[scale=0.4]{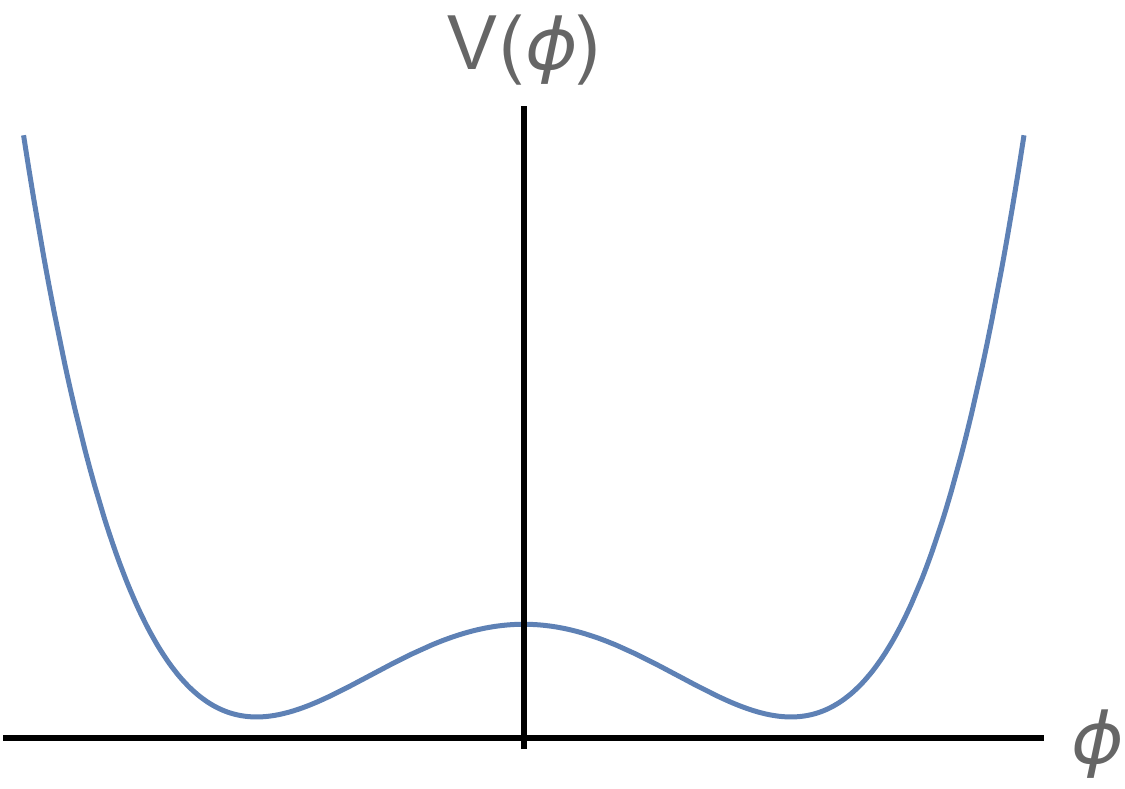}
\qquad
\includegraphics[scale=0.4]{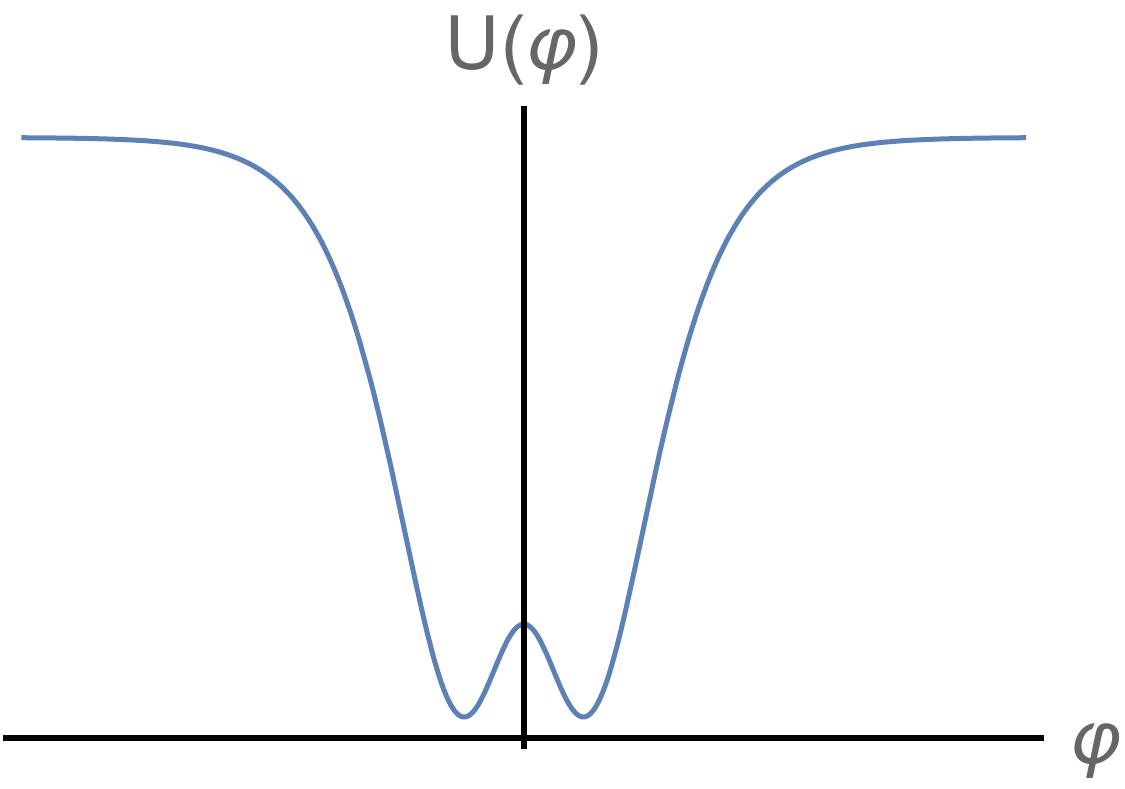}
\qquad
\includegraphics[scale=0.4]{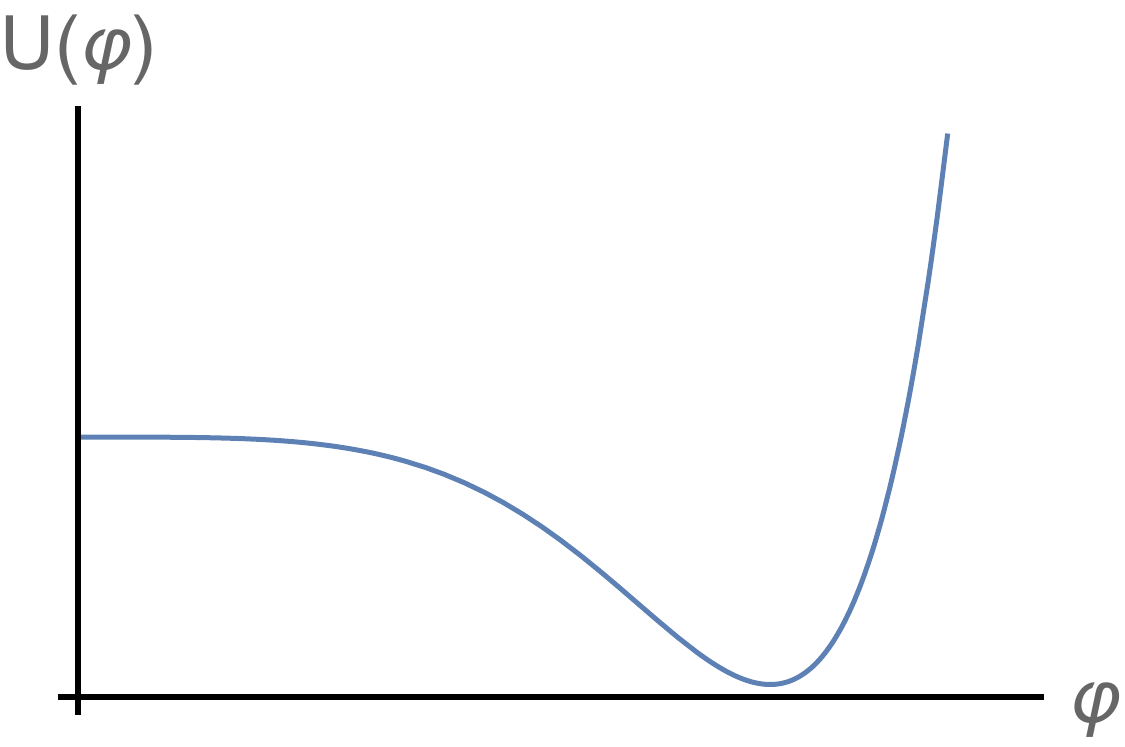}
  \caption{The effect of canonicalising a potential $V(\phi)$ to $U(\varphi)$ with a pole of order $p\ge 2$ and $p<2$, respectively. In the first case, the canonical potential features an plateau that stretches all the way to infinity, and in the second case, a finite plateau close to $\varphi = 0$. Note that for $p<2$, $\varphi<0$ is not shown, since the non-canonical field approaching to the pole corresponds to the canonical field approaching 0.}
  \label{fig:stretch}
\end{figure}

Up until this point, our analysis uses the original field $\phi$. We may arrive at the same results by redefining the inflaton using \eqref{multifieldredef} such that the kinetic term becomes canonical:
\begin{align}
\label{canoninfl}
\varphi =
\begin{cases}
 \frac{2\sqrt{\alpha_p} |\phi -\phi_p| ^{1-\frac{p}{2}}}{p-2} 	& (p \ne 2),\\
\sqrt{\alpha_p}  \ln  |\phi -  \phi_p|  					& (p = 2).
\end{cases}
\end{align} 
Depending on the order of the pole,  we observe that the canonical potential $U(\varphi) = V(\phi(\varphi))$ becomes
\begin{align}
\label{Ucan}
U(\varphi) =
\begin{cases}
 V(0)  \left[1+  \left(\frac{p-2}{2\sqrt{a_p} \varphi}\right)^{-2/(p-2)}\right] 	& (p \ne 2),\\
V(0) \left[ 1 +  e^{-\varphi/\sqrt{\alpha_p}} \right]		& (p = 2).
\end{cases}
\end{align}
It suffices now to use the standard expressions for $\epsilon$ and $\eta$ (with $k = 1$) in order to arrive at the observables \eqref{observables}. While the results are naturally the same, the attractor behaviour is now evident in a different, perhaps more intuitive way: the potential is ``stretched'' into a plateau as $\varphi \to \infty$. Depending on the order of the pole, this plateau will be finite, corresponding to hilltop inflation (for $p<2$) or infinite, corresponding to ``inverse-hilltop'' inflation (in the case that $p\ge 2$)~\cite{Terada:2016nqg}. This can be seen schematically in Figure~\ref{fig:stretch}.

\begin{figure} 
  \centering
  \hspace{-1em}
\includegraphics[scale=0.50]{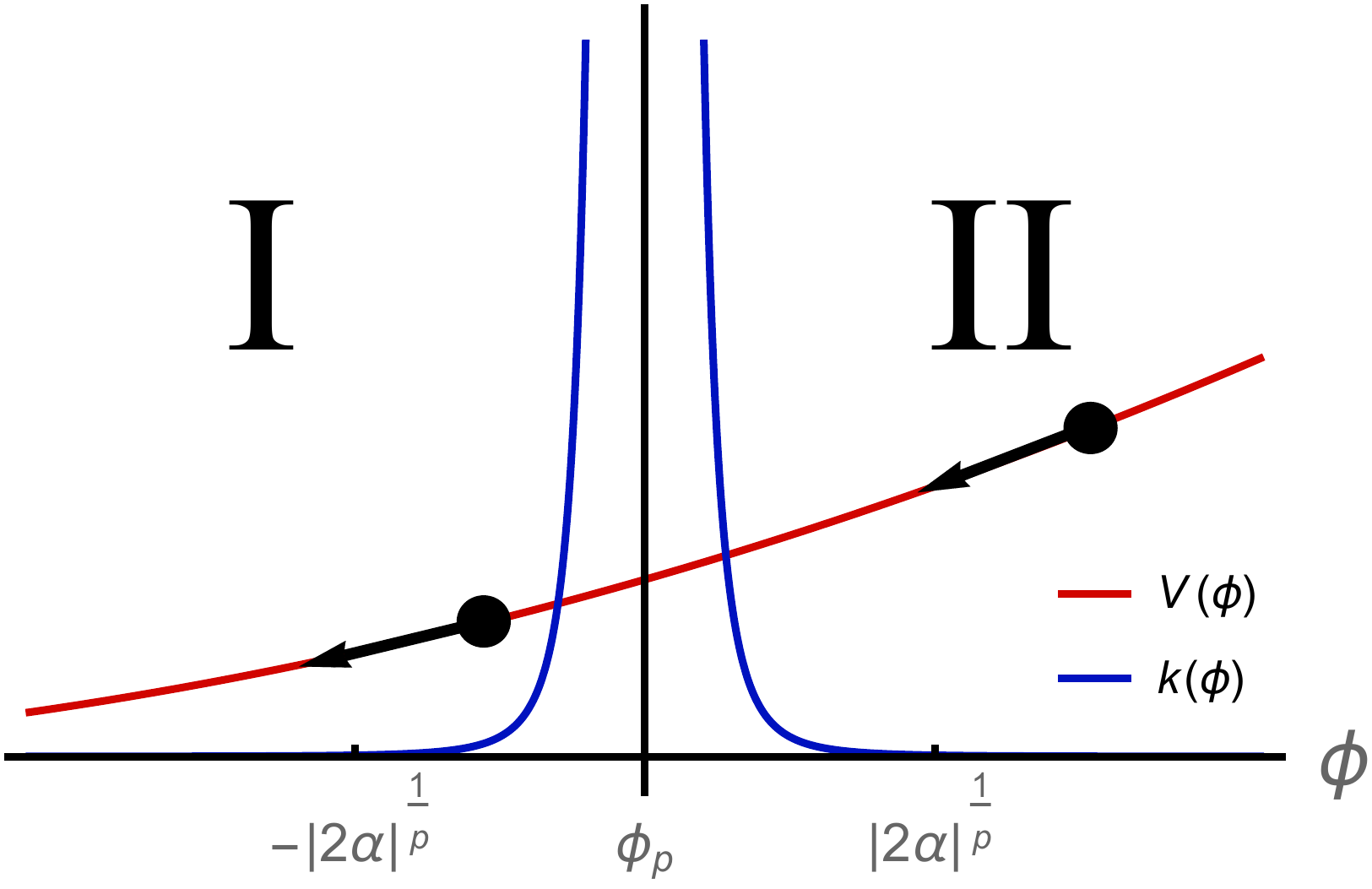}
  \caption{The evolution of a non-canonical field with a singular (positive) kinetic term subject to a monotonic potential. The field has been shifted such that the pole is at the origin. In domain I, the field rolls away from the pole, and when it crosses $\phi = -|2\alpha_p|^{1/p}$, inflation stops. On the contrary, in domain II, late-time acceleration begins once the field crosses the $-|2\alpha_p|^{1/p}$.}
\label{fig:twodomains}
\end{figure}

An important condition in order for the expressions \eqref{observables} for the observables to make sense in the context of inflation is that the gradient of the potential is such that it pushes the field away from the pole, an assumption we have already made but not explained. First, we note that it is the non-canonical (original) potential that determines whether we approach or recede from the pole.\footnote{This is easy to see by analogy to classical mechanics: even if the mass of a particle is time-dependent (which here corresponds to $k(\phi)$ when the fields are homogeneous), the sign of $V'(\phi)$ is the only thing that determines whether the motion is to the left or the right.}
Close to the pole, we can use \eqref{srp} to find that $\epsilon \propto \phi^p$. If the field is pushed towards the pole, the Universe will then enter a period of late-time acceleration. In this case, $N$ is not counted backwards from the end of inflation (as usual), but rather forward from the \emph{start} of the accelerated expansion. In this case, we can see that the longer the Universe remains quasi-de Sitter and $N$ grows, the more it approaches the scale-invariant spectrum. If the pole is crossed by a strictly monotonic potential, one side will support inflation, and the other one late-time (and eternal) acceleration, as shown in Figure~\ref{fig:twodomains}.

We have shown that in general, the shape of the potential close to a pole is irrelevant, in accordance to pole inflation, but choosing which \emph{side} of the pole we choose can strongly affect the predictions of the theory. In particular, it is possible that one side will support early-time acceleration (if the field is pushed away from the pole), and one side will support late-time acceleration (if the field is pulled towards the pole).
This is another indication that when a pole exists in the Lagrangian, we need a second canonical potential in order to recover the original Lagrangian, as discussed in Section~\ref{sttheories}.  We can see this in another way: if we start with an arbitrary potential and attempt to ``decanonicalise'' the inflaton by imposing an arbitrary kinetic term with a pole, we cannot determine what the shape of the non-canonical potential would be on the other side of the pole. 
A single canonical potential is not enough to recover a non-canonical action with a specified kinetic term without some information as to what the non-canonical potential looks on the other side of the pole. Therefore, choosing which side of the pole one starts is not just a matter of initial conditions, but also a matter of choice of which model we use. We will see this explicitly in the following subsection by constructing a model that can give rise to different phenomenology to all orders in $N$.
   
 \subsection{Master models}
   
In Section~\ref{sttheories}, we saw how poles in a Lagrangian lead to multiple canonical potentials. The converse also holds: multiple canonical Lagrangians with different canonical potentials can be ``glued'' together into a ``master Lagrangian'' that corresponds to different single-field models. Consider the following non-canonical Lagrangian
\begin{align}
\label{masterlagrangian}
(-g)^{-1/2} \mathcal{L} = -\frac{1}{2} R + \frac{1}{2\phi^2} \partial^\mu \phi \, \partial_\mu \phi - {\cal V }(\phi) ,
\end{align}
where ${\cal V }(\phi)$ is a potential to be determined later. 
Canonicalising the field leads to the following simple relation between $\varphi$ and $\phi$:
\begin{align}
\varphi=   \ln \frac{1}{|\phi|}.
\end{align}
We remind that we cannot cross the pole at $\phi=0$, but $\varphi$ goes over the entire real line, so the sign of the non-canonical field will depend on whether we inflate to the left or to the right of the pole. Now, consider two distinct canonical potentials $U_-(\varphi)$ and $U_+(\varphi)$. We can ensure that canonicalising the potential $V(\phi)$ on the different sides of the pole will lead to these potentials if
\begin{align}
{\cal V }(\phi) =
\begin{cases}
U_+\left(   \ln \frac{1}{ \phi }\right)  &(\phi>0)
\\
U_-\left(  \ln \left[-\frac{1}{ \phi }\right]\right) & (\phi<0).
\end{cases}
\end{align}
This procedure of ``gluing'' together two completely distinct potentials can be repeated indefinitely, simply by adding more kinetic poles to a model and choosing an appropriate  master non-canonical potential.
\begin{figure} 
  \centering
  \hspace{-1em}
\includegraphics[scale=0.5]{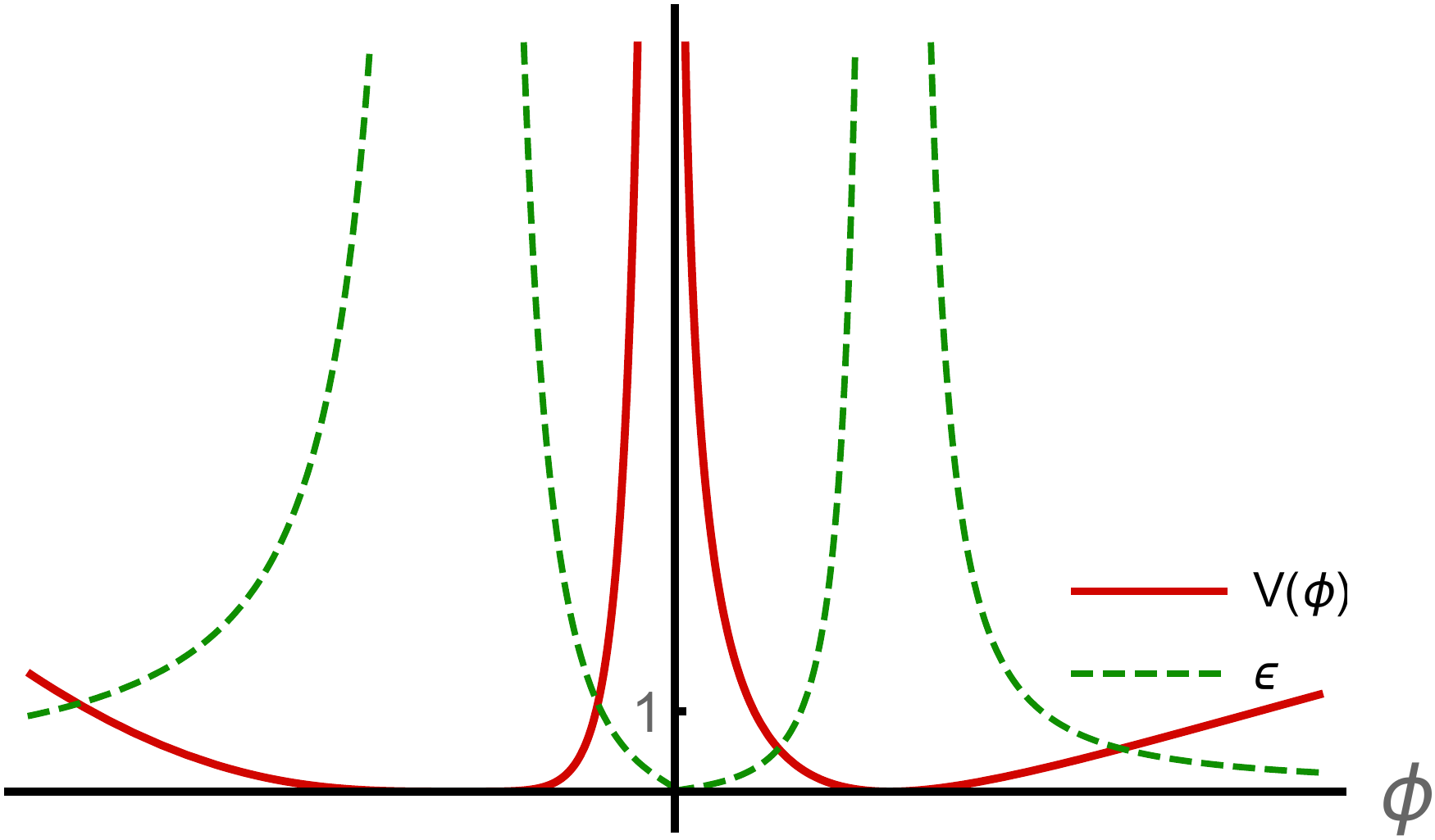}
  \caption{The master non-canonical potential for a glued $\varphi^4$ and $\varphi^2$ theory and the associated slow-roll parameter $\epsilon$. Inflating on the left side of the pole corresponds to inflation with a quartic potential and inflating on the right hand side corresponds to a quadratic potential. Note that, as expected, the slow-roll parameter $\epsilon$ is asymptotic at the two points that correspond to $\varphi = 0$ in the two canonical potentials $U_-(\varphi)$ and $U_+(\varphi)$. Inflation stops as usual at $\epsilon =1$, which occurs twice in each domain.}
  \label{fig:gluing}
\end{figure}

If we take  $U_+ (\varphi)= m^2 \varphi^2$ and $U_- (\varphi)= \lambda \varphi^4$, the potential ${\cal V}(\phi)$ takes on the explicit form
\begin{align}
{\cal V }(\phi) =
\begin{cases}
m^2  \ln \left( \dfrac{1 }{\phi}\right)^2  &(\phi>0)
\\  
\lambda   \ln \left(- \dfrac{1 }{\phi}\right)^4  &(\phi<0).
\end{cases}
\end{align}
This potential features a pole as well, as seen in Fig.~\ref{fig:gluing}. This does not further ``split'' the field space, since the pole in the potential is at the origin, just like the pole in the kinetic term. 

We now check whether the predictions for the master Lagrangian are going to be the same as $\varphi^4$ and $\varphi^2$ 
depending on the side of the pole. To the left of the pole, we use the generalised slow-roll parameters \eqref{srp} to find
\begin{align}
n_\mathcal{R} &= 1-\frac{24}{\ln ^2\left(-\frac{1}{\phi }\right)} ,
&
r &= \frac{128}{\ln ^2\left(-\frac{1}{\phi }\right)}.
\end{align}
For  large $N$ where we can ignore $\varphi_{\rm end}$ where $\epsilon = 1$, and so
\begin{align}
N=
\frac{1}{8} \ln ^2\left(-\frac{1}{\phi }\right).
\end{align}
We may write down the observables for $\phi<0$ in terms of $N$ to first order:
\begin{align}
\label{phi4obs}
n_\mathcal{R} &= 1 - \frac{3}{N} ,
&
r &= \frac{16}{N}.
\end{align}
Similarly, for the right-hand side, we find 
\begin{align}
n_\mathcal{R} &= 1-\frac{8}{\ln ^2\left(-\frac{1}{\phi }\right)} ,
&
r &= \frac{32}{\ln ^2\left(-\frac{1}{\phi }\right)},
\end{align}
and
\begin{align}
N=
\frac{1}{4} \ln ^2\left(-\frac{1}{\phi }\right),
\end{align}
leading to the following expressions for the observables for $\phi>0$:
\begin{align}
\label{phi2obs}
n_\mathcal{R} &= 1-\frac{2}{N} ,
&
r &= \frac{8}{N}.
\end{align}
We observe that, as expected, the observables \eqref{phi4obs} correspond to those for a $\varphi^4$ theory, while the observables \eqref{phi2obs} correspond to those for a $\varphi^2$ theory. This occurs because the potential violates \eqref{limnoneq}; it is not continuous, and it is therefore possible to have entirely different phenomenology depending on the side of the pole.
 
In general, we can ``glue'' as many single-field models as we want together, and we may decompose any model with $n$ kinetic poles into $n+1$ distinct single-field models. We write an action featuring $n$ kinetic poles of order $2$ as:
\begin{align}
(-g)^{-1/2}\mathcal{L} = -\frac{1}{2} R + \left(\frac{1}{2} \sum_{i=1}^{n} \frac{1}{(\phi-\phi_i)^2} \right)^2 (\partial^\mu \phi)(\partial_\mu \phi) - {\cal V }(\phi),
\end{align}
Canonicalising, we find
\begin{align}
\varphi(\phi)= \sum_{i=1}^{n} \frac{1}{\phi_i -\phi}.
\end{align}
This function is not one-to-one over the entire real line, but it is one-to-one within each of the following intervals
\begin{align}
\begin{aligned}
I_1 &= (-\infty, \phi_1),
\\
I_i &= (\phi_{i-1}, \phi_{i}),
\\
I_{n+1} &= (\phi_{n}, \infty),
\end{aligned}
\end{align}
where $i \ne 1, n+1$.
Finally, if the canonical potentials we wish to glue together are $U_1(\varphi), \ldots U_{n+1}(\varphi)$, we find that the master potential is given by the following piecewise function:
\begin{align}
{\cal V} (\phi)  = 
\begin{cases}
 U_1(\varphi(\phi)) & (\phi \in I_1),
\\  
\vdots
 \\  
U_{n+1}  (\varphi(\phi)) &  (\phi \in I_{n+1}).
\end{cases}
\end{align}
This procedure can be performed backwards, which is the case most often encountered in the literature. Of course, this may prove cumbersome, as $\varphi(\phi)$ is not necessarily easy to invert, but as long as we are careful to clarify which interval we restrict ourselves to, it is always possible to do so.

We have seen how the presence of poles in a Lagrangian can hide the existence of radically different models that might lead to different predictions for the inflationary observables if the derivative of the inflationary potential is not continuous. However, this is not the end of the story. Even if the observables converge close to the poles, the late-time behaviour of singular Lagrangians might be starkly different depending on which side of the pole we are on. In the next section, we will examine  $\alpha$-attractors, and show that going beyond the usual confines of the poles, it is possible to realise an observationally sound inflation scenario that nonetheless features a novel behaviour, as the field manages to reach the end of the field space.

\section{$\alpha$-attractor models within and beyond the poles}
\label{alphaattractors}

We have studied how the presence of poles in a  Lagrangian results in a  one-to-many mapping in Lagrangian space. In order to illuminate this feature with a more realistic example, we turn our attention to the well-studied case of $\alpha$-attractors \cite{Kallosh:2013wya,Kallosh:2013xya}. From a purely phenomenological point of view, the Lagrangian for $\alpha$-attractors is given by
 \begin{align}
\label{actionJalpha}
(-g)^{-1/2}\mathcal{L} =   - \frac{R}{2} + \frac{1 }{2} \frac{(\partial_\mu \phi )(\partial^\mu \phi)}{\left(1-\frac{\phi^2}{6\alpha}\right)^2} - V ( \phi)  .
\end{align}
This Lagrangian features two poles at $\phi = \pm \sqrt{6\alpha}$. 
In general, if one treats Lagrangians of the type~\eqref{actionJalpha} as a phenomenological toy model that simply reproduces the predictions of superconformal models inspired from Poincar\'{e} disk hyperbolic geometry~\cite{Carrasco:2015rva}, then considering what happens beyond the poles does not hold much meaning. However, $\alpha$-attractors can arise in other contexts, where the domains beyond the poles are not excluded \emph{a priori}. Moreover, viewing a model with poles as-is, it behooves us to examine how the phenomenology will be affected by choosing a particular interval for the non-canonical field. 

In this section, we will study $\alpha$-attractors both within and beyond the poles. We will first study their potential-independent features, and then move on to specialise to an exponential potential. For this particular case, we will find that three distinct scenarios are realised, one of which corresponds to the well-known quintessential inflation, and one to eternal late-time acceleration.  The final domain features the most novel scenario, in which the scalar field, after driving inflation, becomes infinitely light as it reaches the boundary of the field space.

\subsection{Domain-dependent early- and late-time acceleration}

We have seen that if we have a model with a singular kinetic term, we must study each domain defined by the poles separately. In order for the Universe to accelerate within a given interval, there must be at least one subinterval for which $\epsilon(\phi) < 1$. 
Furthermore, in order to have a transition between accelerating and non-accelerating evolution, the gradient of the potential must push the field \emph{towards} or \emph{away from} $\phi_{\rm end}$ (defined through $\epsilon(\phi) = 1$). The first case corresponds to late-time (eternal) acceleration, whereas the second case corresponds to early-time acceleration, or inflation (which eventually stops, and therefore does not suffer from the graceful exit problem).
As we mentioned earlier, in general the field will roll down the potential, regardless of the presence of the kinetic term. This happens as long as there are no tachyonic modes i.e. the kinetic term remains positive. Therefore, the potential will determine whether the pole is approached or not.

\begin{figure}
\centering
\includegraphics[scale=0.30]{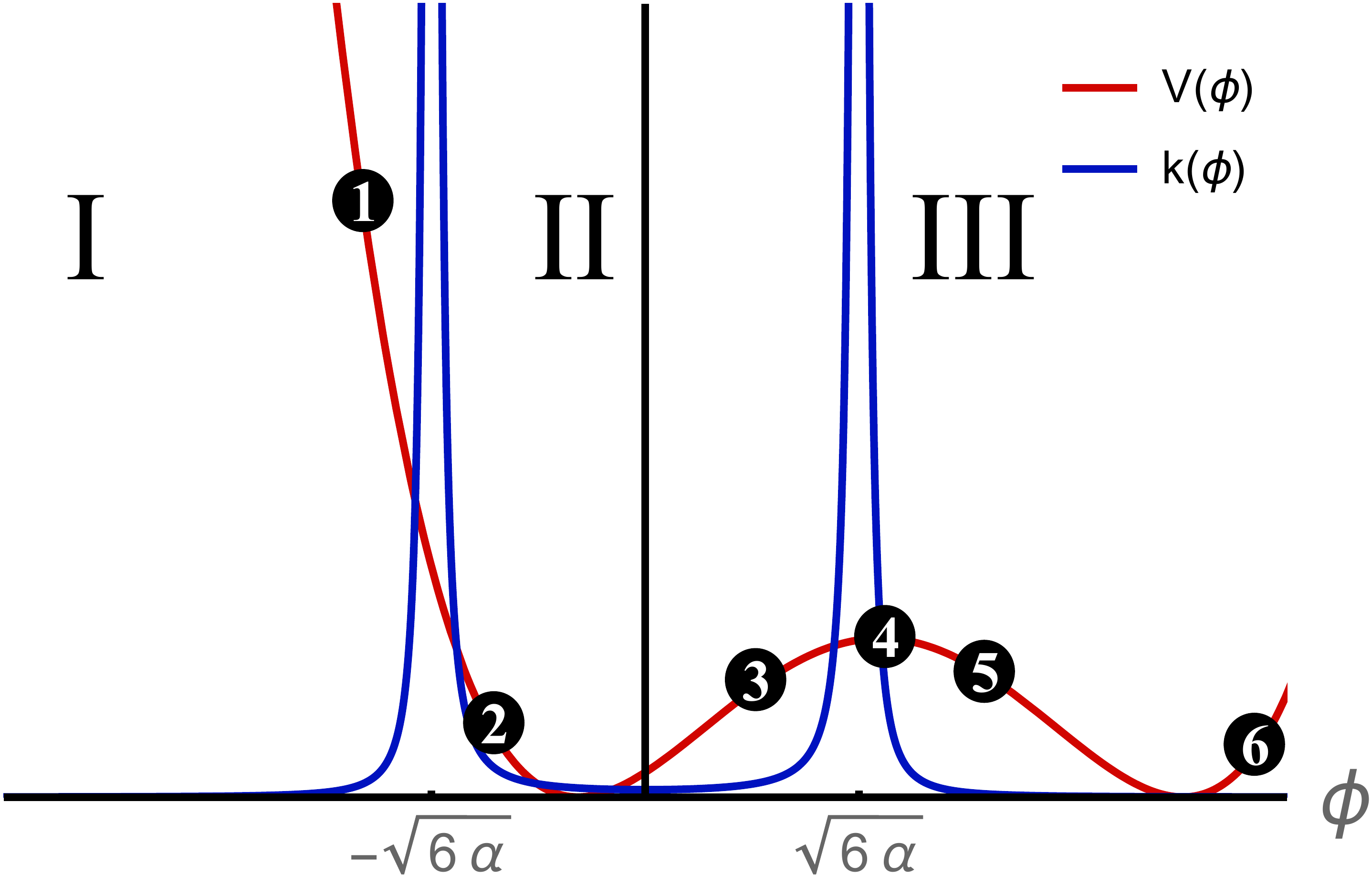}

\includegraphics[scale=0.25]{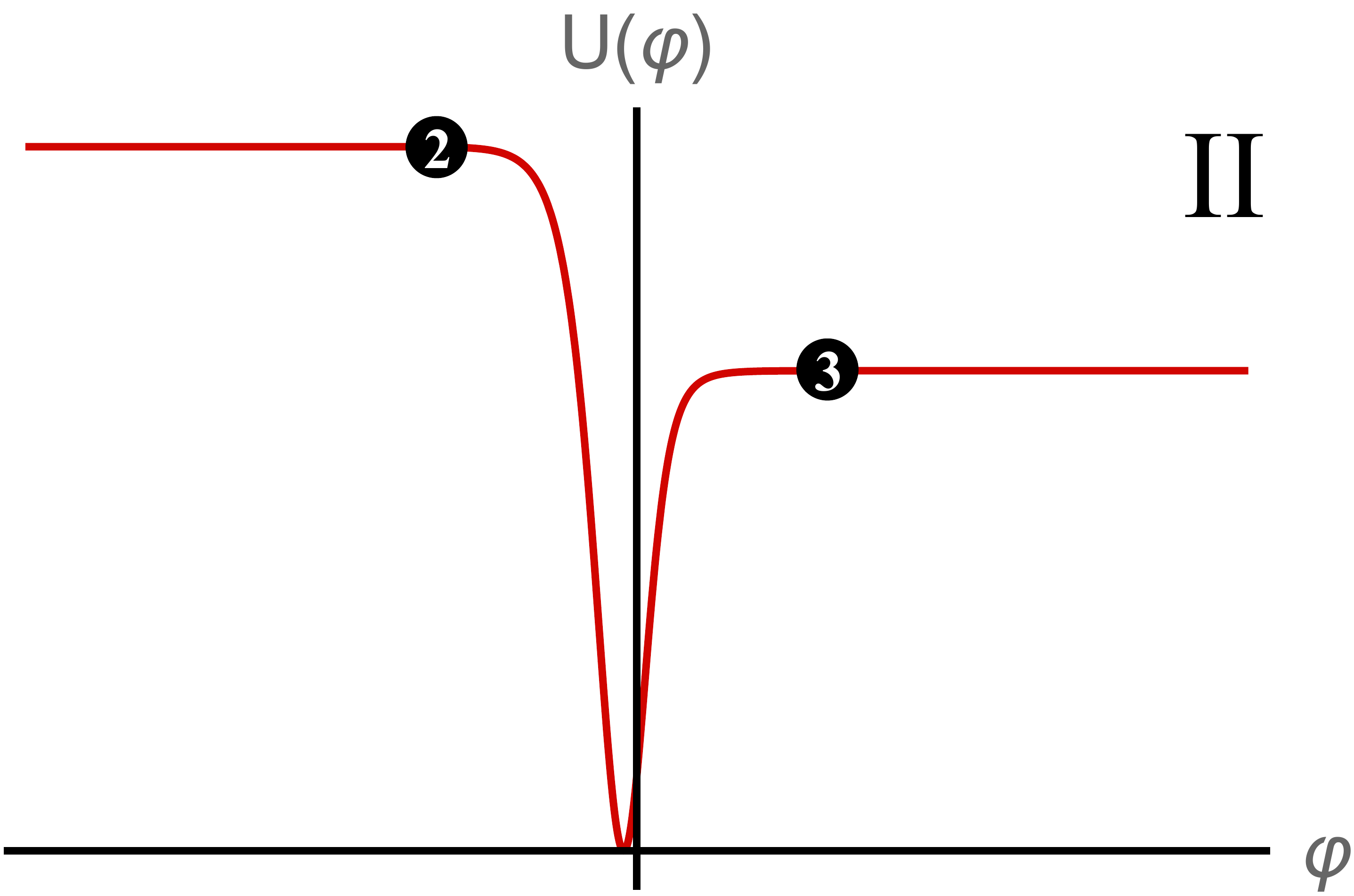}
\includegraphics[scale=0.25]{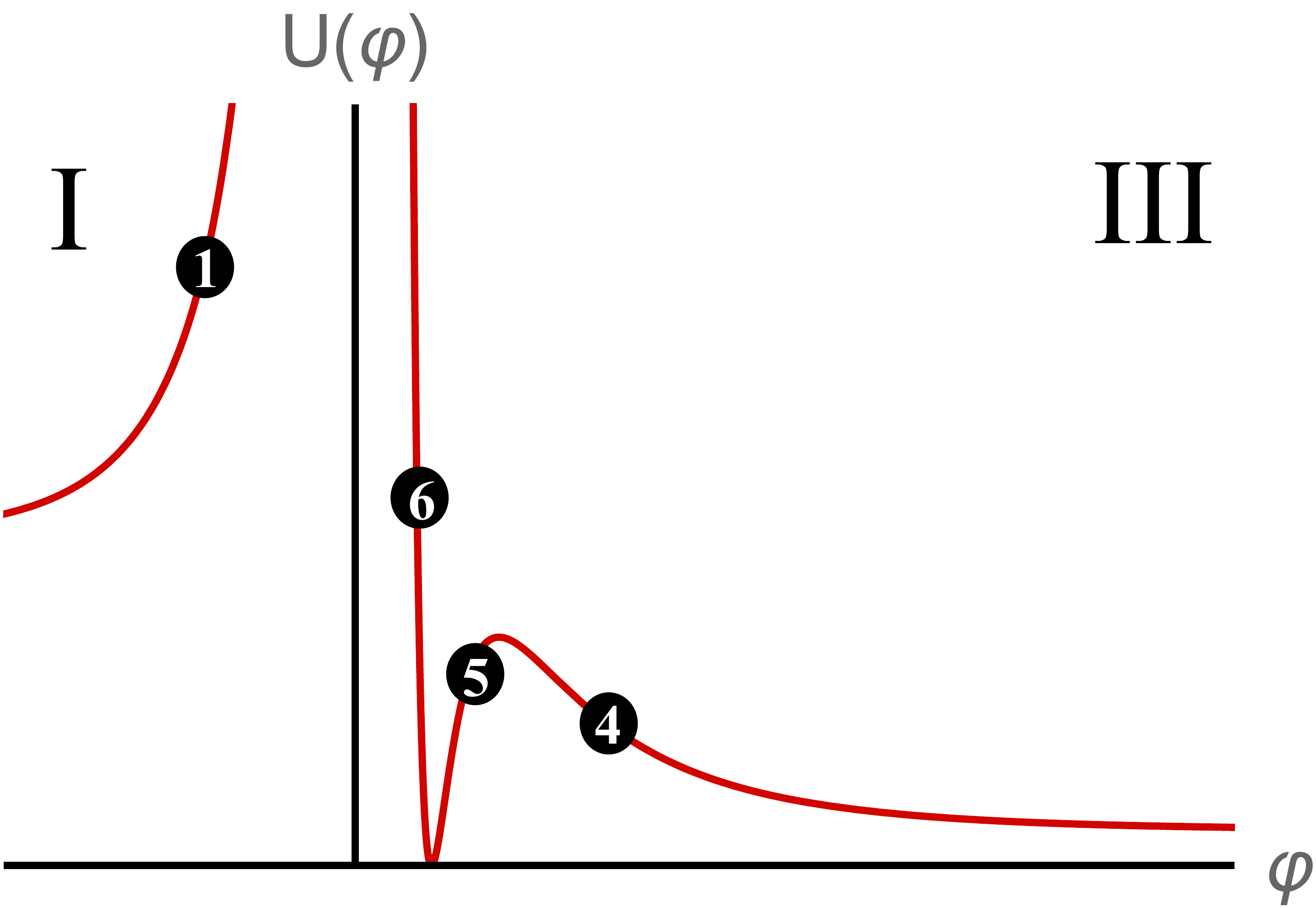}
  \caption{Canonicalised potentials for an $\alpha$-attractor model with an arbitrary potential $V(\phi)$. Domain II is within the poles whereas domains I and III are outside the poles (left and right, respectively). Note that in all domains, the poles correspond to $\varphi = \pm \infty$. The numbers show the correspondence between the canonical and non-canonical field.}
\label{fig:canpot}
\end{figure}

We now canonicalise the field for $\alpha$-attractors. In general, we find that it is given by
\begin{align}
\label{varphigen}
\varphi = \sqrt{\frac{3\alpha}{2}} \left( \ln \left|\frac{ \phi + \sqrt{6\alpha} }{ \phi - \sqrt{6\alpha}}\right| -   \ln \left| \frac{ \phi_0 + \sqrt{6\alpha} }{ \phi_0 - \sqrt{6\alpha}} \right| \right).
\end{align}
Crucially, the value of $\phi_0$, which corresponds to which value for $\phi$ we associate $\varphi = 0$ in \eqref{formsol}, is arbitrary. This means that which interval we choose $\phi_0$ to be in determines which canonical domain we are in: $|\phi_0|<\sqrt{6\alpha}$ within the poles and $|\phi_0|>\sqrt{6\alpha}$ beyond the poles. Within the poles, we usually choose $\phi_0 = 0$, leading to
\begin{align}
\varphi_{\rm in}= \sqrt{6\alpha}\tanh^{-1}\frac{\phi}{\sqrt{6\alpha}}.
\end{align}
In this case, the poles at $\phi = \pm\sqrt{-6\alpha}$ correspond to $\varphi  = \pm \sqrt{6\alpha}$. Outside of the poles, we can choose any point $|\phi_0|>\sqrt{6\alpha}$, and so for simplicity, we let $|\phi_0| \to \infty$. This leads to the following expression for the canonical field:
\begin{align}
\label{phiout}
\varphi_{\rm out}= \sqrt{6\alpha}\coth^{-1}\frac{\phi}{\sqrt{6\alpha}}.
\end{align}
Inverting $\varphi_{\rm in}$ and~$\varphi_{\rm out}$, we may write the following canonicalised potentials that arise within and outside of the poles when canonicalising the original potential~$V(\phi)$:
\begin{align}
\label{Uin}
U_{\rm in} (\varphi) &= V\left( \sqrt{6\alpha}\tanh \frac{\varphi}{\sqrt{6\alpha}}\right),
\\
\label{Uout}
U_{\rm out}(\varphi)  &=  V\left( \sqrt{6\alpha}\coth \frac{\varphi}{\sqrt{6\alpha}}\right),
\end{align}
We note that $\varphi$ runs only over $(0,\infty)$ as $\phi>\sqrt{6\alpha}$, and similarly over $(-\infty,0)$ as $\phi<-\sqrt{6\alpha}$. This fact means that we actually require only two canonical potentials to fully specify the model, even if there are three domains due to the presence of two poles.  The canonical potentials for the two domains are sketched  in Figure~\ref{fig:canpot} for an arbitrary non-canonical potential $V(\phi)$.

We can now study the domain between the poles, which is the case most often encountered. We focus on the positive pole by expanding \eqref{Uin} for large $\varphi$ as
\begin{align}
\label{Uingiven}
\begin{aligned}
U_{\rm in}(\varphi) 
&= V\left( \sqrt{6\alpha}- \sqrt{6\alpha} e^{-2\varphi/\sqrt{6a}}\right)
\\
&\approx V'(\sqrt{6\alpha}) \left[ -v ( \sqrt{6\alpha}) -  e^{-2\varphi/\sqrt{6a}}\right],
\end{aligned}
\end{align}
where $v$ is defined in \eqref{defv}. For early-time acceleration, the potential must push the field away from the pole, which occurs for the initial conditions $2$ and $3$ in Figure~\ref{fig:canpot}. If we restrict ourselves to large positive values of~$\varphi$, then $v$ must be negative.
Furthermore, We observe that, no matter what the shape of the original potential $V(\phi)$, the canonical potential $U_{\rm in}(\varphi)$ will plateau to $V(\pm \sqrt{6\alpha})$ as $\varphi\to\pm\infty$. Therefore, the height of the original (non-canonical) potential will be unimportant for determining $n_\mathcal{R}$ and $r$, as outlined in the previous section. This allows us to recast the potential in the following form by stretching and shifting $\varphi$:
\begin{align}
\label{Uexpapprox}
U_{\rm in}(\widehat\varphi) 
&\propto  1-  e^{-\beta\widehat\varphi},
\end{align}
where $\beta = 2/\sqrt{6\alpha}$. This is possible since the dimensionless observables (ignoring the total strength of the spectrum) are not going to depend on $v$. This potential corresponds to Starobinsky inflation \cite{Terada:2016nqg}. The observables can be calculated to be
\begin{align}
\begin{aligned}
n_\mathcal{R} &= \frac{1 -\beta ^2-2 \left(\beta ^2+1\right) e^{\beta  \widehat\varphi }+e^{2 \beta  \widehat\varphi } }{\left(e^{\beta  \widehat\varphi }-1\right)^2},
\\
r &= \frac{8 \beta ^2}{\left(e^{\beta  \widehat\varphi }-1\right)^2}.
\end{aligned}
\end{align}
The number of $e$-foldings is found to be
\begin{align}
 N  =   \frac{1}{\beta^2}  (  e^{ \beta\widehat\varphi}- e^{ \beta\widehat\varphi_{\rm end}})- \frac{1}{\beta}(\widehat\varphi -\widehat\varphi_{\rm end}  ),
\end{align}
where 
$\varphi_{\rm end} = \beta^{-1}\ln  \frac{1}{2} \left(\sqrt{2} \beta +2\right) $. 
Therefore, for a large number of $e$-foldings (which occurs as the field~$\widehat \varphi \to \infty$), we can arrive at the standard expressions for the observables:
\begin{align}
\begin{aligned}
n_\mathcal{R} &=\frac{4 \beta ^2 (N-2) N+4 \sqrt{2} \beta  (N-1)-10}{\left(2 \beta  N+\sqrt{2}\right)^2},
\\
r &= \frac{32}{\left(2 \beta  N+\sqrt{2}\right)^2}.
\end{aligned}
\end{align}
For large values of $\beta$, which is to say small values of $\alpha$ (small enough that $\alpha \ll 2/3 N \approx 30$), we arrive at the usual results:
\begin{align}
n_\mathcal{R} &=
1 - \frac{2}{N}  +\frac{\sqrt{3\alpha}  }{N^2} -\frac{3 \alpha  (3 N+1)}{2 N^3}+ {\cal O}(\alpha^{3/2}),
\\
r &=  \frac{12 \alpha }{N^2}-\frac{12 \sqrt{3} \alpha ^{3/2}}{N^3} + {\cal O}(\alpha^{2}).
\end{align}
These expressions may diverge to second order if we use the actual form of the potential instead of the expansion \eqref{Uexpapprox}. We may additionally scale the potential to match amplitude of the scalar power spectrum to the observed value of $P_R = 2.24 \times 10^{-9}$ at $N  \approx 60$ (often given as $(V/\epsilon)^{1/4} = 0.027)$, leading to a phenomenologically sound model. This is a straightforward example of the attractor behaviour studied in the previous section. 

We now turn our attention to the phenomenology of the domain outside of the poles. Away from the pole (which is to say for small $\varphi$), the potential \eqref{Uout} becomes
\begin{align}
\label{canpot3}
U_{\rm out}(\varphi) = V\left(\frac{6\alpha  }{\varphi }\right).
\end{align}
We see that outside of the poles, the canonical potential will be singular if the original potential $V(\phi)$ is not singular, as demonstrated in \eqref{fig:canpot}. Close to the pole (for large $\varphi$), the potential is
\begin{align}
\label{canpot4}
U_{\rm out}(\varphi) = V\left(\sqrt{6\alpha } + \sqrt{6\alpha} e^{-\sqrt{\frac{2}{3\alpha}} \varphi } \right).
\end{align}
Therefore, we can stretch and expand the field in a similar way to \eqref{Uexpapprox}. The potential then becomes proportional to $1+e^{-\beta\widehat\varphi}$. Once again, the overall shape of the canonicalised potential close to the poles is independent of the original potential: it tapers off to a constant value at infinity, leading to attractor behavior when calculating the observables. However, if the canonical potential is monotonic as it crosses the pole, inflation cannot be realised simultaneously on both sides of the pole.  We can see this schematically in Figure~\ref{fig:canpot}: early-time acceleration occurs for initial condition 5 and 6 (although the attractor behaviour is subdued for the latter since the field is far away from the pole, even if it approaches), and late-time acceleration occurs for initial conditions 1 and~4 as the canonical field rolls away to infinity (which corresponds to the non-canonical field approaching the pole). 
 
We have seen that for $\alpha$-attractors, three distinct domains arise which may have diverging predictions. In the next subsection, we will examine $\alpha$-attractors with an exponential potential, which lead to quintessential inflation between the poles. We will show that in the domains beyond the poles, a novel behaviour arises in which the field reaches the end of the field space in finite time.

\subsection{Quintessential inflation within the poles}

Having studied $\alpha$-attractors in general, we now turn our attention to quintessential inflation~\cite{Peebles:1998qn}, which features both early- and late- time acceleration within the poles. Quintessential inflation can be quite straightforwardly realised in the context of $\alpha$-attractors~\cite{Akrami:2017cir,Dimopoulos:2017zvq,Dimopoulos:2017tud}, and is specified by \eqref{actionJalpha} with $V(\phi) = V_0 e^{-\kappa \phi}$. The canonical potential within the poles acquires the following form:
\begin{align}
\label{Uinexact}
U_{\rm in}(\varphi) &= V_0 \exp\left[-\kappa\sqrt{6\alpha} \tanh \left(\frac{\varphi }{\sqrt{6\alpha}}\right)  \right].
\end{align}
Therefore, as discussed in the previous subsection, there are three domains: $\sqrt{6\alpha}<\phi$, $-\sqrt{6\alpha}<\phi<\sqrt{6\alpha} $, and $\phi<\sqrt{6\alpha}$, shown in~Figure~\ref{fig:canpot}. We will examine these separately.
  \begin{figure} 
  \centering
\includegraphics[scale=0.5]{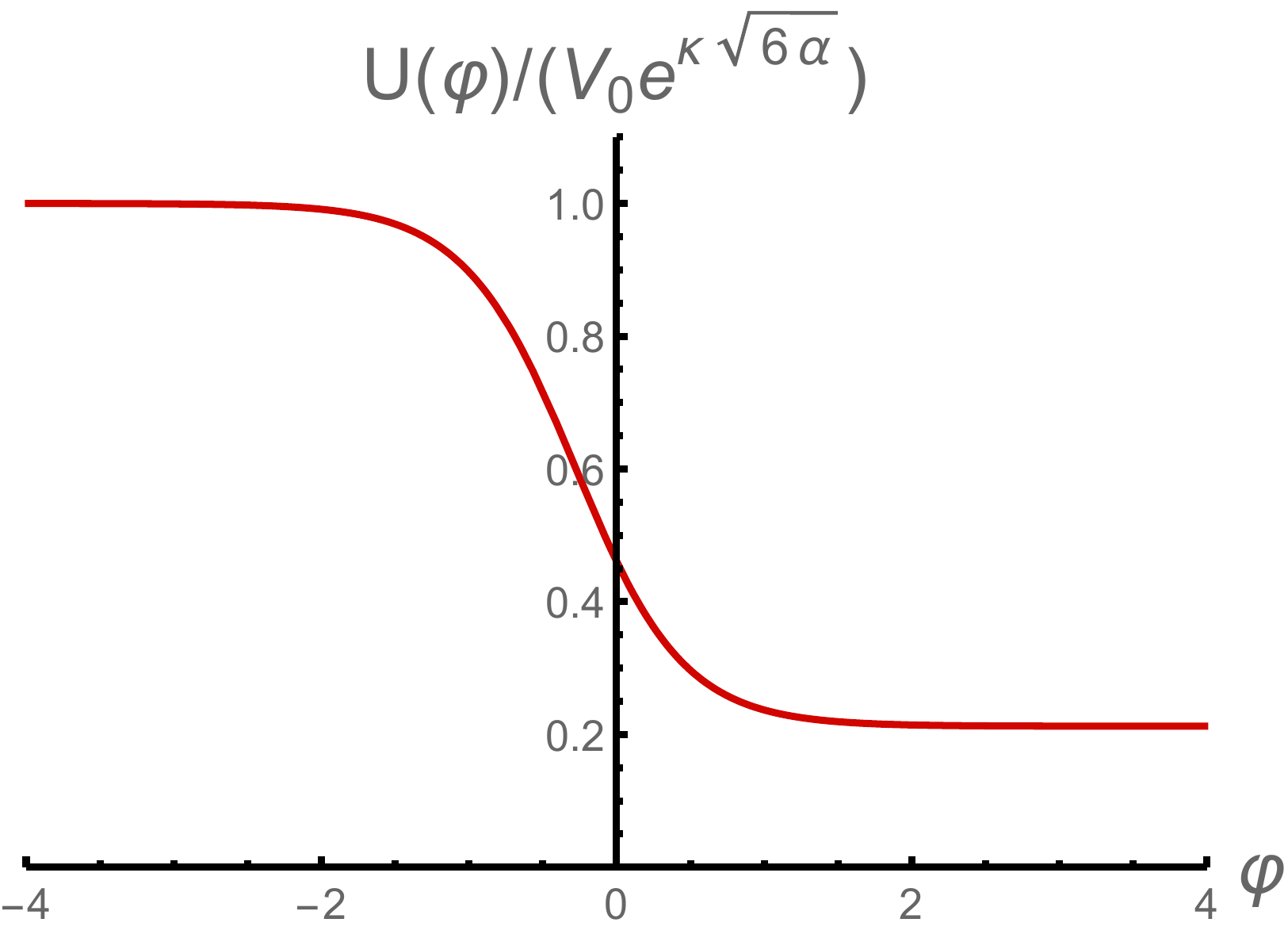}
  \caption{The canonicalised potential (within the poles) for an $\alpha$-attractor model with $V(\phi) \propto e^{-\kappa\phi}$. In order to achieve the correct value for the cosmological constant, we are required to offset the potential, but when it comes to inflation, this is not necessary.}
\label{fig:quintpot1}
\end{figure}

We first look at the potential within the poles. For large negative values of $\varphi$, the potential is given by
\begin{align}
\label{Uinapprox}
U_{\rm in}(\varphi) \approx V_0 \exp\left[-\kappa\sqrt{6\alpha} \left( 1 - e^{ 2\varphi/\sqrt{6\alpha}} \right)  \right].
 \end{align}
Usually, an offset of the order of the electroweak scale is required in order to realise observationally sound quintessence, but as we focus on inflation, we will not be performing this step. Using the potential \eqref{Uinapprox}, the observables are calculated to be
\begin{align}
\begin{aligned}
n_\mathcal{R} &=1 
+4 \kappa \sqrt{\frac{2}{3\alpha}}  e^{  {\sqrt{\frac{2}{3\alpha}} \varphi } }
-4 \kappa ^2 e^{ {2 \sqrt{\frac{2}{3\alpha}} \varphi } },
\\
r &=
32 \kappa ^2 e^{ {2 \sqrt{\frac{2}{3\alpha}} \varphi }  }.
\end{aligned}
\end{align}
We can find where the end of inflation occurs by using the exact expression for the potential given in~\eqref{Uinexact}:
\begin{align}
\label{}
\epsilon = \frac{1}{2} \kappa ^2 \text{sech}^4\left(\frac{\varphi }{\sqrt{6\alpha}  }\right).
\end{align}
We therefore find that $\kappa^2>2$ is required for inflation to end. Working with the approximate potential \eqref{Uinapprox}, we find that the end of inflation happens at
\begin{align}
\label{eoinf2}
\varphi_{\rm end} = -\frac{1}{2}\sqrt{\frac{3\alpha}{2}}  \ln \left(2 \kappa ^2\right).
\end{align}
Therefore, the number of $e$-foldings is found to be
\begin{align}
N = \frac{1}{2 \kappa }\sqrt{\frac{3 \alpha }{2}} \big(  e^{\sqrt{\frac{2}{3 \alpha }} \varphi_{\rm end}  } -  e^{\sqrt{\frac{2}{3 \alpha }} \varphi }\big).
\end{align}
We are therefore able to write the observables as follows:
 \begin{align}
 \begin{aligned}
n_\mathcal{R} &= 
\frac{-3 \alpha +4 \sqrt{3\alpha}   +4 \left(\sqrt{3\alpha}  +2\right) N + 4 N^2}{\left(\sqrt{3\alpha}  +2 N\right)^2},
\\
r &=
\frac{48 \alpha }{\left(\sqrt{3\alpha}  +2 N\right)^2}.
\end{aligned}
\end{align}
Expanding for small $\alpha$, the observables become
 \begin{align}
 \begin{aligned}
n_\mathcal{R} &= 
1-\frac{2}{N}
-\frac{\sqrt{3\alpha}    }{   N^2}
-\frac{3 \alpha  (N-1)}{2 N^3} + \mathcal{O}(\alpha^{3/2}),
\\
r &=
\frac{12 \alpha }{N^2}
-
\frac{12 \sqrt{3} \alpha ^{3/2}  }{   N^3} + \mathcal{O}(\alpha^2).
\end{aligned}
\end{align}
We can see that the potential $U_{\rm in}(\varphi)$ supports inflation for large negative values of $\varphi$. The relation between $\kappa$ and $\alpha$ can be found by matching to the strength of the scalar spectrum for a large number of $e$-foldings:
\begin{align}
 \label{PRmatch}
 \frac{4 V_0 N^2 e^{-\sqrt{6\alpha}  \kappa }}{3 \alpha } \approx (0.027)^4. 
\end{align}
In order to achieve $r<0.1$, we have $\alpha \lesssim 30$. The natural value of $\alpha$ is usually close to 1, and so we find that $\kappa \approx 10$ for $V_0 \approx 1$.

After examining the inflationary predictions for this model, we briefly review the evolution of inflation after inflation ends. The following overview closely follows~\cite{Dimopoulos:2017zvq}. When $\varphi_{\rm end}$ is reached, the energy of the field is mostly kinetic as $\epsilon$ quickly surpasses unity. This period is referred to as \emph{kination}, with an equation of motion $\ddot\varphi + 3 H\dot\varphi = 0$, leading to
 \begin{align}
\label{varphiend}
\varphi -  \varphi_{\rm end} =    \frac{2}{3} \ln (t/t_{\rm end})  .
\end{align}
This equation is model-independent, although $\varphi$ must be the canonical field (otherwise the kination equation of motion would not be as simple). During kination, the scale factor goes as $a\propto t^{1/3}$. If $\Omega_\gamma^{\rm end}$ is the radiation density parameter at the end of inflation, reheating begins at $t_{\rm reh} = (\Omega_\gamma^{\rm end})^{-3/2} t_{\rm end}$ (the point at which radiation takes over).
Since the eventual evolution of the field is not oscillatory, reheating in such models is achieved via alternative mechanisms, such as instant preheating~\cite{Felder:1998vq}, curvaton reheating~\cite{Feng:2002nb}, or gravitational reheating~\cite{Ford:1986sy, Chun:2009yu}.  

Substituting the time of reheating into \eqref{varphiend}, we find that the total distance traversed by the field $\varphi$ is
 \begin{align}
\Delta \varphi_{\rm reh}=  -\sqrt{\frac{3}{2}} \ln \Omega_\gamma^{\rm end} .
\end{align}
After kination ends, the evolution of the field is still dominated by the potential. However, in radiation domination, the equation of motion is
 \begin{align}
\dot\varphi  = \frac{2}{3} \frac{\sqrt{t_{\rm reh}}}{t^{3/2}}.
\end{align}
Therefore, after integrating, we find that the field will freeze after traversing the following distance after the end of inflation:
 \begin{align}
\label{fieldexc}
\Delta \varphi_{\rm F} = \sqrt{\frac{3}{2}} \Big(1 - \frac{3}{2}\ln \Omega_\gamma^{\rm end}\Big).
\end{align}
After this occurs, the potential can give rise to a positive cosmological constant. Indeed, we observe that the model features a quintessential tail for large positive values of~$\varphi$ (as long as a proper offset is added):
 \begin{align}
U_{\rm in}(\varphi) &\approx V_0 \exp\left[-\kappa\sqrt{6\alpha} 
\left(1 - e^{ {-2\varphi /\sqrt{6\alpha}} }\right) \right].
\end{align}
As the potential approaches the potential energy of the attractor (for dominant or subdominant quintessence) and transients die down, the field ``unfreezes'' and begins to evolve once again. In any case, it can lead to accelerated expansion, which may be eternal or transient, depending on the steepness of the tail. The details are beyond the scope of this paper:  the interested reader is referred to~\cite{Dimopoulos:2017zvq,Dimopoulos:2017tud}. What is important is that the same field which drives inflation can be used to drive quintessence as well.
 
\subsection{Quintessential inflation beyond the poles and the field space boundary}

\begin{figure} 
  \centering
\includegraphics[scale=0.5]{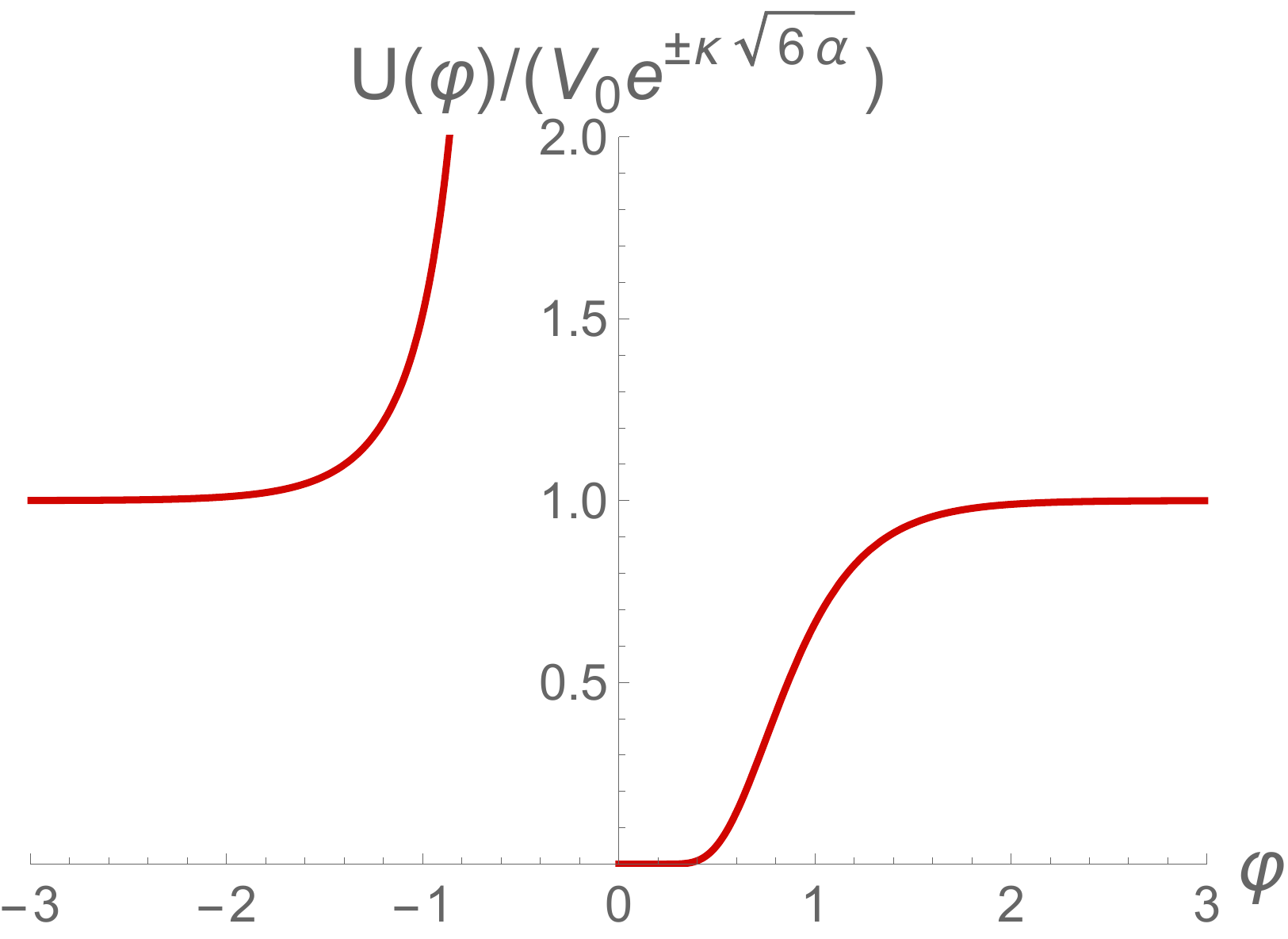}
  \caption{The canonicalised potential outside the poles for an $\alpha$-attractor model with $V(\phi) \propto e^{-\kappa\phi}$. The sign for the normalisation depends on the domain: $-$ for $\phi>\sqrt{6\alpha}$ and $+$ for $\phi<-\sqrt{6\alpha}$.}
\label{fig:quintpotout}
\end{figure}

We now turn to examine the domain beyond the poles. The canonical potential in this case is given by
\begin{align}
\label{Uoutexact}
U_{\rm out}(\varphi) &= V_0 \exp\left[-\kappa\sqrt{6\alpha}
\coth\left(\frac{\varphi}{\sqrt{6\alpha}}   \right)
\right].
\end{align}
This potential is sketched in Figure~\ref{fig:quintpotout}. The domain $\varphi<0$ cannot support inflation: acceleration never ends as $\phi$  is pushed away from the pole. Therefore, we focus on $\phi>\sqrt{6\alpha}$.

We now turn our attention at the phenomenology of this scenario as the field moves away from the pole. Looking at the potential outside the poles given \eqref{Uoutexact} and calculating~$\epsilon$, we find that the end of inflation occurs at
\begin{align}
\varphi_{\rm end} &= \sqrt{\frac{3\alpha}{2}}  \ln \left[  \sqrt{2\kappa ^2+2\sqrt{2} \kappa }+\sqrt{2} \kappa +1\right]
\\
&\approx  \frac{1}{2}\sqrt{\frac{3\alpha}{2}}  \ln  (2\kappa),
\end{align}
where the approximate value holds for $\kappa \gg 1/\sqrt{2}$. Unlike quintessential inflation within the poles, $\kappa > 1/\sqrt{2}$ is not required for graceful exit. For large values of $\varphi$, the canonical potential is approximated as:
\begin{align}
\label{btpquint}
U_{\rm out}(\varphi) \approx \exp \left[ - 2 \kappa \sqrt{6 \alpha }\,  e^{ - \varphi/\sqrt{6\alpha}  } 
\coth\left( \frac{\varphi}{\sqrt{6\alpha}} \right)
\right].
\end{align}
This potential is similar to the one within the poles for large values of $\varphi$, and so we expect that it can give rise to phenomenologically supported inflation. We now calculate the observables beyond the poles are found to be
\begin{align}
\begin{aligned}
n_\mathcal{R}
&=
1
- 4 \sqrt{\frac{2}{3\alpha}} \kappa   e^{ -2\varphi/\sqrt{6\alpha}  }
-4 \kappa ^2   e^{ -4\varphi/\sqrt{6\alpha}  },
\\
r&= 32 \kappa ^2   e^{ -4\varphi/\sqrt{6\alpha}  }.
\end{aligned}
\end{align}
The number of $e$-foldings can be found as
\begin{align}
N = 
\frac{1}{2 \kappa } \sqrt{\frac{3\alpha}{2}}   \left(   e^{ -2\varphi/\sqrt{6\alpha}  } -   e^{ -2\varphi_{\rm end}/\sqrt{6\alpha}  }\right).
\end{align}
Inverting the above, we finally find the expressions for the observables in terms of~$N$:
\begin{align}
\begin{aligned}
n_\mathcal{R}
&=
\frac{3 \left[4 N^2+4 \left(\sqrt{3\alpha }  -2\right) N-3 \alpha -4 \sqrt{3\alpha}  \right]}{\left(3 \sqrt{\alpha }+2 \sqrt{3} N\right)^2},
\\
r&= \frac{144 \alpha }{\left(3 \sqrt{\alpha }+2 \sqrt{3} N\right)^2}.
\end{aligned}
\end{align}
Once again, we expand for small $\alpha$ to find
\begin{align}
\begin{aligned}
n_\mathcal{R}
&=
 1-\frac{2}{N} +\frac{\sqrt{3\alpha}  }{N^2}-\frac{3 \alpha  (N+1)}{2 N^3} + {\cal O}(\alpha^{3/2}),
\\
r&=\frac{12 \alpha }{N^2}-\frac{12 \sqrt{3} \alpha ^{3/2}}{N^3} + {\cal O}(\alpha^2).
\end{aligned}
\end{align}
These expressions for the observables are similar to those of quintessential inflation. Once again, by appropriately scaling $V_0$, it is possible to match to the strength of the scalar power spectrum using an analogous relation to \eqref{PRmatch}, which is 
\begin{align}
\label{normout}
  \frac{4 N^2 V_0 }{3\alpha   } \exp \left[ -    4  \kappa^2   N  \right] \approx (0.027)^4.
\end{align}
Therefore, after selecting $\alpha$, we are left once again with a relation between $V_0$ and $\kappa$.

The canonical potential beyond the poles as seen in Figure~\ref{fig:quintpotout} has a unique feature: as the canonical field rolls down to 0, this corresponds to $\phi$ rolling off to infinity in finite time. This is no cause for alarm: there is no physical singularity associated with this behaviour. While the field excursion $\Delta \phi$ is infinite when expressed in terms of the non-canonical $\phi$, but it is important to remember that the field excursion is a chart-dependent quantity. What matters is the field-space distance travelled, which is given by
\begin{align}
\label{fielddist}
\Delta \varphi= \int_{\phi_0}^\infty \frac{d\phi}{ 1-\frac{\phi^2}{6\alpha} } < \infty,
\end{align}
which coincides with the canonical field excursion, hence our choice of notation $\Delta \varphi$. Note that, as usual, it is impossible to reach the poles because the field must traverse an infinite field distance to do so according to \eqref{fielddist}. However, reaching $\phi=\pm \infty$ is possible. Within the poles, the field lives on a projective line, whose points at infinity ($\varphi = \pm \infty$) correspond to the poles. Beyond the poles however, the field lives on a projective \emph{ray}: one pole is indeed located at infinity ($\varphi = \infty$ when in domain III in Figure~\ref{fig:canpot}), but the field space ends at $\varphi = 0$.

\begin{figure} 
  \centering
\includegraphics[scale=0.45]{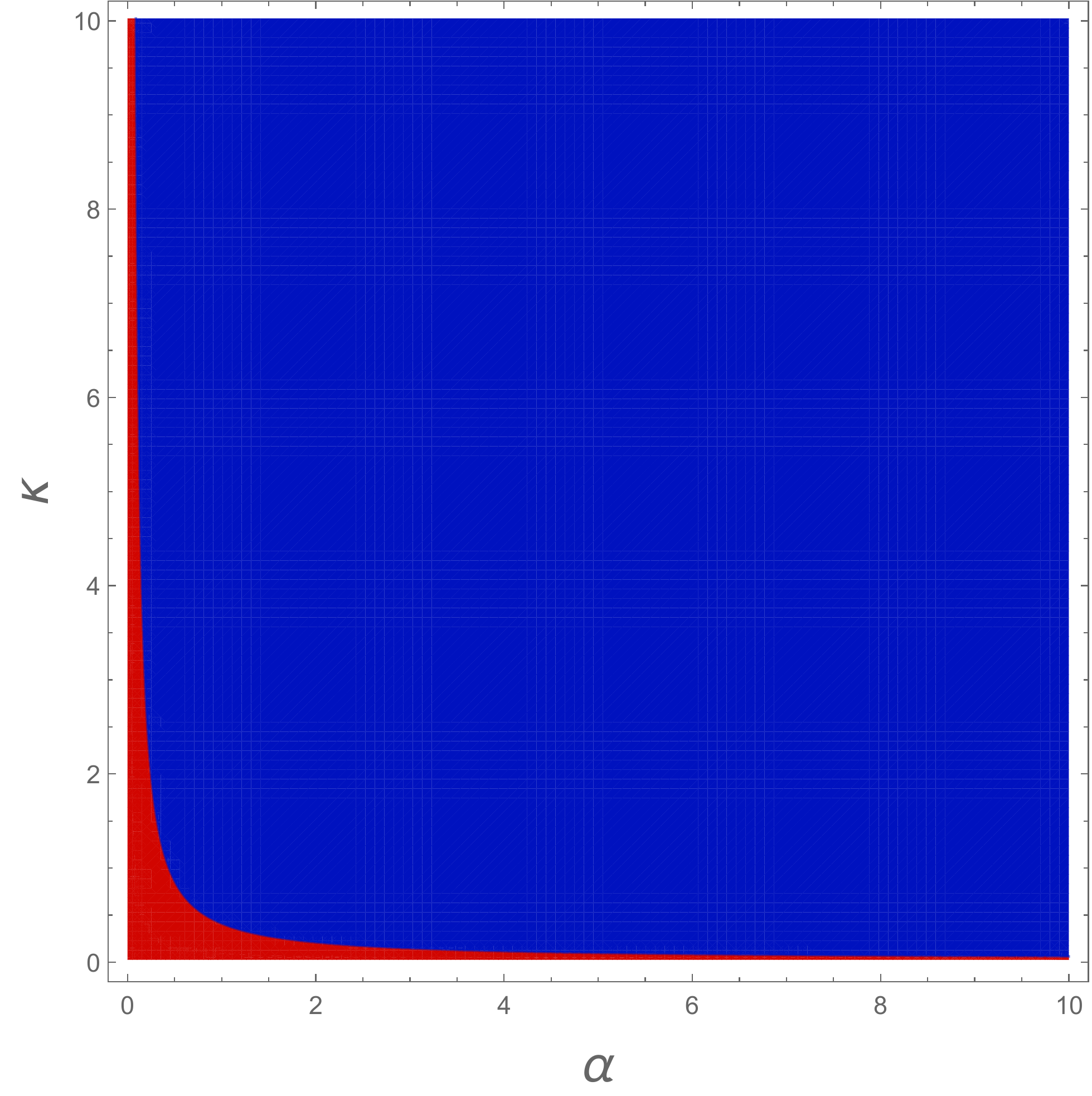}
  \caption{Parameter space for $\alpha$ and $\kappa$ under the assumption of instant preheating. The freezing of the scalar field is completed in the blue region and interrupted in the red region as the field reaches the boundary of the field space.}
\label{fig:alphakappa}
\end{figure}

As discussed in the previous subsection, the field will eventually stop evolving and freeze at a value $\varphi_F = \varphi_{\rm end} - \Delta \varphi_F $, where the distance it traverses after the end of inflation is given in \eqref{fieldexc}. At that point, the field has lost most of its kinetic energy as well as its potential energy. However, eventually it will approach the attractor solution again, and begin to evolve. As long as the field has frozen before reaching the boundary of the field space, however,  it will asymptotically approach reach a potential energy and kinetic energy of zero, and will essentially completely decouple. We now determine the parameter ranges that enable the field to freeze before it reaches the edge of the field space. Using the total excursion of the field after \eqref{fieldexc} for $  \Omega_\gamma^{\rm end}  = 1$ (corresponding to instant preheating\footnote{The largest possible excursion occurs for gravitational reheating, but we focus on instant preheating for simplicity, which can occurs quite generically for single-field and multifield models induced by non-minimal couplings, roughly 2-2.5 $e$-foldings after the end of inflation \cite{Nguyen:2019kbm}.}), the freezing of the field will occur before reaching the boundary if
 \begin{align}
\label{fieldexcconstr}
\sqrt{\alpha}  \ln \left[  \sqrt{2\kappa ^2+2\sqrt{2} \kappa }+\sqrt{2} \kappa +1\right] > 1.
\end{align}
If this relation is satisfied, then the Hubble drag is going to be enough to prevent the field $\varphi$ reaching~0 before it freezes. The region in parameter space for $\alpha$ and $\kappa$ satisfying the above relation is shown  in Figure~\ref{fig:alphakappa}. We observe that, for most natural values of $\alpha$ and $\kappa$, the freezing does complete before the field reaches the boundary of the field space.
  
Just after the evolution of the field is arrested, the potential will approximately take on the value
\begin{align}
\label{Uoutapprox}
U_{\rm out}(\varphi) &\approx V_0 \exp\left[- \frac{   6\alpha\kappa  }{ \varphi_F  } \right].
\end{align}
Using \eqref{normout}, we may find (using $\alpha \approx  1$)
\begin{align}
V_0 \approx \ 1.11 \times 10^{-10}  \times  e^{60  \kappa ^2}.
\end{align}
The value of the frozen field for the same value of $\alpha = 1$ is
\begin{align}
\varphi_F = \sqrt{\frac{3 }{2}} \left(  \ln \left[  \sqrt{2\kappa ^2+2\sqrt{2} \kappa }+\sqrt{2} \kappa +1\right] - 1\right).
\end{align}
The above two equalities can be used to determine $\kappa$ using the coincidence requirement$U_{\rm out} (\varphi_F) \approx 2.89 \times 10^{-122}$:
\begin{align}
 U_{\rm out} (\varphi_F) = 1.11 \times 10^{-10} \exp\left[ 60  \kappa ^2- \frac{   6 \kappa  }{ \sqrt{\frac{3 }{2}} \left(  \ln \left[  \sqrt{2\kappa ^2+2\sqrt{2} \kappa }+\sqrt{2} \kappa +1\right] - 1\right)} \right], 
\end{align}
where we have assumed $N \approx 60$ (the exact value will depend on how close reheating is to instant preheating). 
The coincidence requirement therefore determines $\kappa \approx 0.2515828$ (a large degree of accuracy is required; this is because the value of the potential is very sensitive to $\kappa$). We can also perform a consistency check: this value barely prevents the field from reaching the end of the field space with $\varphi_F \approx 0.00579$. We have therefore illustrated that it is possible to have quintessence beyond the poles that remains consistent with observations, without the requirement of an offset, as is usual in traditional quintessential inflation. While this expansion may be eternal, which can cause issues such as the future horizon problem in string theory \cite{Hellerman:2001yi,Fischler:2001yj}, it is still consistent with observations.

There is one question left to be answered: what happens if the field does reach the end of the field space? Discussing this is largely academic, as we expect the field to freeze before it does so in order to obtain a positive cosmological constant. However, we may trace the root of this issue to the fact that theory described by the non-canonical Lagrangian \eqref{actionJalpha} is, in a sense,  incomplete. By itself, it cannot tell us what happens beyond the boundary. As such, we must impose a boundary condition on the point at infinity in order to determine what happens after it is reached. The most minimal assumption is perhaps the most unphysical: a first-order phase transition which sees the kinetic energy of the inflaton instantaneously vanish as it reaches the boundary and remains there. This is a true case of ``vanishing inflation'': the field is stationary and therefore couples to nothing. However, if we are not willing to make this assumption, we must analytically extend the manifold by appending another projective ray to it, only this time pointing to the other direction. We have already defined the canonical field such that it vanishes at infinity in \eqref{phiout}, allowing us to write the original theory as:
\begin{align}
(-g)^{-1/2}\mathcal{L} = \left[ -\frac{R}{2} + \frac{k(\phi(\varphi))}{2} \; \partial_\mu \phi(\varphi)  \,  \partial^\mu \phi(\varphi)  - V(\phi(\varphi)) \right]. \;
\end{align}
The physical meaning of the value of $\varphi$ is the field space distance from the boundary of the manifold, and as such, it must be positive. The simplest way to analytically continue the manifold is therefore to allow $\varphi$ to take on negative values: in this case, the potential can take on any functional form. Therefore, in the extended theory, the potential is
\begin{align}
V (\phi (\varphi) ) =  
\begin{cases}
U_{\rm out}(\varphi) &  \varphi>0 \;, 
\\
U_{\rm extended} (\varphi) &  \varphi<0 \; .
\end{cases}
\end{align}
We are free to choose  $U_{\rm extended} (\varphi)$ as we wish, perhaps subject to continuity constraints (although this is not required classically, it will affect the quantisation of the theory). We could choose to make the theory truly massless beyond the boundary (by setting $U_{\rm extended} (\varphi) = 0$), or we could mirror the already known potential as $U_{\rm extended} (-\varphi) = U_{\rm out}(\varphi)$. Nonetheless, it is clear that this is a choice we must make. This is a consequence of the incompleteness of the theory: we must extend it ``by hand'' by specifying what the potential is beyond the boundary. Note that we do not need to specify the kinetic term, since it can be always set to unity in the same domain. While it may vanish at the boundary (since $\phi = 0$), this occurs only at a single point, and as such the set of points where the kinetic term is canonical is almost the entire space.
 
In this section, we have reviewed the phenomenology of $\alpha$-attractors both within and beyond the poles equipped with an exponential potential. We have studied the common setup between the poles, leading to quintessential inflation. However, when examining the model outside of the poles, we find that we can also achieve quintessence, as long as the field freezes before it reaches the boundary of the (semi-) finite field space. We also study the extension of the field space that is required when the evolution of the field is arrested as it reaches the boundary, concluding that the field space must be equipped with an analytically continued potential in order to determine the eventual fate of the inflaton.

\section{Beyond the poles in multifield theories}
\label{modelwalls}
 
Our discussion up to this point has focused mostly on single-field models. We have seen that a given Lagrangian can act as a ``sum'' of different models when it features poles.
However, multifield inflation may also feature kinetic terms with singularities. It is usually assumed that such singularities may not be crossed, but this is not necessarily always the case: we know from general relativity, for instance, that the Schwarzschild metric may feature a coordinate singularity at the Schwarzschild radius that can be eliminated by appropriate choice of coordinates. We therefore expect that there may exist singularites in field space that arise due to a ``bad'' parametrisation of the scalar fields. These are nothing more than coordinate singularities, as opposed to physical singularities. These are defiend to be $(n-1)$-dimensional surfaces that cannot be crossed by the fields subject to the equations of motion.. In this section, we will demonstrate that, in analogy with single-field models, Lagrangians with true singularities can be decomposed in terms of singularity-free Lagrangians. For convenience, we will focus on two-field models, but our results can be extended for more than two fields.

\subsection{Model walls in two-field models}
\label{subsec:modelwallstwofield}

 We begin by considering a two-field Lagrangian as a special case of \eqref{minnoncan}, with two scalar fields  $\phi$ and~$\chi$:
\begin{align}
\label{diagonalL}
 a^{-3}\mathcal{L} =  -\frac{ R}{2} + \frac{1}{2} \frac{ \dot \phi^2}{P(\phi,\chi)} + \frac{1 }{2}\frac{ \dot \chi^2 }{Q(\phi,\chi)}- V(\phi,\chi).
\end{align}
Here, we have assumed (without loss of generality) that the field-space metric $G_{AB}$ is diagonal with $G_{\phi\chi} = 0$ (this is always possible to do, since the metric is symmetric). We have ignored spatial derivatives as we assume that the inflaton is homogeneous, and we have assumed that the metric is FLRW, i.e. $g_{\mu\nu} = {\rm diag}(1,-a^2,-a^2,-a^2)$ in Cartesian coordinates. If $P(\phi,\chi)  = 0$ or $Q(\phi,\chi)  = 0$, the kinetic term shoots off to infinity.  As opposed to single-field inflation, where the kinetic terms become singular at certain points, this occurs along $(n-1)$-dimensional \emph{singular surfaces} embedded in the $n$-dimensional field space.  
  
After we have identified singular surfaces for a Lagrangian, the question that arises is whether they can be crossed, which will determine if they are physical or not. For single-field inflation, the answer is straightforward: ``curves'' in this case are poles, and the canonical field cannot cross them because that would always require moving an infinite field-space distance in finite time (which essentially requires infinite kinetic energy). This indicates that the pole can never be actually be reached (much less crossed). However, when there are additional fields, this is not immediately obvious. In order to be sure that a singular curve cannot be crossed, we could solve the equations of motion, but this is not necessary. Field space, as opposed to spacetime, does have a notion of ``absolute time'' in the sense that the affine coordinate $t$ used to parametrise trajectories as $\varphi^A = \varphi^A (t)$ is not itself a coordinate of the manifold. As such, if we can show that there is no finite path connecting the two sides of a singular curve, then we can rest assured that this is a real singularity, since no on-shell path exists connecting the two paths.
Note that we specify \emph{on-shell path} as opposed to \emph{geodesic} since there is a conservative force and a Hubble drag term: in their absence, on-shell paths do correspond exactly to geodesics. 

How do we know whether an on-shell path crossing a singular curve exists? It might actually be easier to show that such a curve does not exist. Consider the field-space line element for the Lagrangian \eqref{diagonalL} when there is only one singular curve:
\begin{align}
\label{flse}
d\sigma^2 =   {d\phi^2}  +  \, \frac{d\chi^2}{Q(\phi,\chi)}.
\end{align}
The singular curve, $Q(\phi,\chi) = 0$. We note that, if the line element is always non-negative (in order to avoid tachyonic modes), crossing the singularity will make the line element diverge. This is a subtle difference between spacetime and field space: in the Schwarzschild metric, for instance, as we cross the horizon, the Schwarzschild time coordinate $t$ goes to infinity in such a way as to cancel out the singularity in the proper time:
\begin{align}
\label{schwarzschild}
d\tau^{2} =  \left(1-\frac{2GM}{r} \right) dt^2 - \left(1-\frac{2GM}{r}\right)^{-1}dr^2- r^2 d\Omega^2.
\end{align}
Therefore, crossing the horizon does occur in finite proper time for an observer falling into the black hole. However, this behaviour cannot occur for the field-space metric \eqref{flse}. As noted above, on-shell paths in field space are parametrised with respect to an affine coordinate for which generically $\phi(\infty) = \infty$. Furthermore, the signature of the metric is always positive for the fields $\phi$ and $\chi$ (to avoid tachyonic modes) and so there can be no cancellation of infinities. As a result, the only way to ensure that $d\sigma^2 < \infty$ is to set $d\chi = 0$ while crossing the singular curve. This is always possible \emph{unless} $Q(\phi,\chi) = Q(\chi)$. In this case, the singular curve is going to be a horizontal line in the $\{\phi,\chi\}$ chart, and crossing it is impossible, since $d\chi = 0$ is parallel to the curve. 

The above condition is sufficient for a singular curve to be a real singularity. While it is highly chart dependent, it can be useful as a rule of thumb. Let us say we observe that for a particular field denoted $\phi^1$ (without loss of generality), there is a term in the field-space line element $(d\phi^1)^2/Q(\phi^1)$. If all terms are positive, we can be certain there is no finite path crossing the singular curve. This may not always be evident from a specific parametrisation, but if we can indeed transform to a chart where this is the case, we can be certain the singularity is real. This is similar to the spacetime case: in the Schwarzschild metric \eqref{schwarzschild}, the two patches $r>r_s$ and $r<r_s$ appear disconnected, but $d\tau^2<\infty$ for on-shell solutions. These patches are therefore \emph{not} actually disconnected, since there is some path for which $d\tau^2$ is finite (and this path must be a spacetime geodesic). The geometric meaning is simple: if there are parts of the spacetime or field-space manifold that are infinitely apart, they cannot communicate. In this sense, real singularity curves in field space can be thought of as \emph{model walls}, much like poles in single-field inflation; they separate truly distinct models.

This convenience in determining at least when a singularity is real (by looking at the line element) is afforded to us because of the nature of the field space. In spacetime, we must generally solve the equations of motion in order to discriminate between real and coordinate singularities (by examining whether the \emph{proper time} goes to infinity as we try to approach the singularity). In field space, it is often easy to see that the element $d\sigma^2$ will go to infinity no matter what path (on-shell or off-shell) we follow, thanks to the fact that $d\phi^A <\infty$ for $dt<\infty$ and the positive signature of the metric. In general, however, if we cannot find a chart in which the above condition is satisfied, we do need to solve the equations of motion to determine the nature of the singularity. In the end, whether they can be crossed on-shell is the only discriminating characteristic between real and coordinate singularities.

An example of a coordinate singularity can be seen in the Eisenhart lift formalism for field theories \cite{Finn:2018cfs}, which may be used to replicate the effects of a potential by constructing a purely kinetic Lagrangian with an additional scalar degree of freedom. Indeed, consider the following Lagrangians:
\begin{align}
\label{L1}
{\cal L}\ &=\   \frac{k_{ij}}{2}(\bm{\phi})\;\dot{\phi}^i\dot{\phi}^j\: -  V(\phi) \;,
\\
\label{L2}
{\cal L}\ &=\   \frac{1}{2} k_{ij}(\bm{\phi})\;\dot{\phi}^i\dot{\phi}^j\: +\:   \frac{1}{2} \frac{\dot{\chi}^2 }{V(\bm{\phi})} \;,
\end{align}
We observe that the singularity $V(\bm \phi)$ can be crossed with $d\chi = 0$, and so there is some path that takes us across it. We must hence solve the equations of motion to see if such a path can also be on-shell. Indeed, we find that varying the second Lagrangian with respect to~$\chi$ and substituting the resulting equation of motion in the equations of motion for $\phi^i$ will yield the equations of motion for the first Lagrangian. Even if \eqref{L2} features a singularity where $V({\bm \phi}) = 0$, we find that trajectories may still cross this curve. It is as if $\dot\chi^2$ and $V(\phi)$ ``conspire'' to approach zero in such a way that their ratio is not indeterminate, ensuring that this is a coordinate singularity.
 
 \subsection{Multifield $\alpha$-attractors and isocurvature perturbations}
 
In the previous subsection, we  distinguished between true and apparent singularities in multifield models. We are now ready to examine exactly how  multifield Lagrangians featuring singular curves can be thought of as a union of multiple Lagrangians, much like single-field Lagrangians featuring poles are unions of canonical Lagrangians. As a concrete example, we consider the following two-field $\alpha$-attractor inspired Lagrangian:
\begin{align}
\label{2fieldlag}
(-g)^{-1/2} \mathcal{L} \; =    \;  -\frac{1}{2} R  \;  +  \;  \frac{1}{2  }\frac{ \partial^\mu \phi \, \partial_\mu \phi  }{ [ 1-  (\phi^2 + \chi^2)/r^2 ]^2} \; +  \;  \frac{1}{2 } \frac{\partial^\mu \chi \, \partial_\mu \chi}{[ 1-  (\phi^2 + \chi^2)/r^2 ]^2}   \;   - \;  {  V } (\phi,\chi)  \; .
\end{align}
Looking at the kinetic term for $\phi$, we note that there is a circle in field space on which there is a singularity, given by $\phi^2 + \chi^2 = r^2$. The field space in this model is curved, and so it is not possible to simultaneously canonicalise both $\phi$ and $\chi$. We now set out to ``decompose'' this model in a similar manner to single-field Lagrangians in Section~\ref{singlefieldpoles} by finding two separate coordinate charts which eliminate the poles from the Lagrangian. 

Our first step is to write the Lagrangian in terms of polar coordinates $\theta = \tan^{-1} (\chi/\phi)$ and $\rho = \sqrt{\phi^2 +\chi^2}$ as
\begin{align}
(-g)^{-1/2} \mathcal{L} =   -\frac{1}{2} R + \frac{1}{2  }\frac{ \partial^\mu \rho \, \partial_\mu \rho  }{ (  1  -\rho^2/ r^2 )^2} + \frac{\rho^2}{2 } \frac{\partial^\mu \theta \, \partial_\mu \theta }{(1  -\rho^2/ r^2 )^2} - { U} (\rho,\theta).
\end{align}
The ``radial'' potential is given by
\begin{align}
 { U} (\rho,\theta) = V(\rho \cos\theta, \rho \sin \theta ).
\end{align}
Since the kinetic term for $\rho$ has no $\theta$ dependence, we observe that this is indeed a real singularity, as per our prescription in the previous subsection. We will now show how this singularity can act as a model wall, giving rise to two distinct multifield models with distinct geometry, topology, and even entropy generation. 

The origin of the two distinct models, in analogy to $\alpha$-attractors, is that the solution to the following equation (that yields a pole-free Lagrangian) has two domains:
\begin{align}
\frac{d\widehat \rho}{d \rho} = \frac{1}{1  -\rho^2/ r^2}.
\end{align}
Inside the disk $(\rho < r)$, we find that the canonicalised field is given by
\begin{align}
\widehat \rho =  r \tanh ^{-1}\left(\frac{\rho }{r}\right).
\end{align}
The field $\widehat \rho$ runs over the entire real line, while $\rho<r$, which indicates that $\widehat\rho$ can be used to parametrise the entire disc. However, this chart is not enough to parametrise the entire manifold. Instead, outside the disc where $\rho>r$, we find
\begin{align}
\widehat\rho  = \frac{r}{2}\left( \ln \frac{ \rho + r }{ \rho - r} -   \ln \frac{ \rho_0 + r }{ \rho_0 - r} \right),
\end{align}
where $\rho_0$ is an arbitrary circle we wish to identify with $\widehat \rho_{\rm out}= 0$, similarly to the single-field attractor case.  This arbitrary circle can be pushed to infinity much like in \eqref{phiout}. We therefore find that the Lagrangian \eqref{2fieldlag} can be decomposed into two Lagrangians without poles. Inside the disk, the  Lagrangian is
\begin{align}
\label{sfree1}
(-g)^{-1/2} \mathcal{L} =   -\frac{1}{2} R + \frac{1}{2  }  \partial^\mu \widehat \rho \, \partial_\mu \widehat\rho    + \frac{1}{2 } \frac{\sinh ^2(  \widehat\rho / r)}{4 r^2}  \partial^\mu \theta \, \partial_\mu \theta   - { \widehat U}_{\rm in} (  \widehat\rho ,\theta).
\end{align}
Outside the disk, the Lagrangian with $\rho_0$ taken to be arbitrarily large is identical to the above, with the exception of the potential:
\begin{align}
\label{sfree2}
(-g)^{-1/2} \mathcal{L} =   -\frac{1}{2} R + \frac{1}{2  }  \partial^\mu \widehat \rho \, \partial_\mu \widehat\rho    + \frac{1}{2 } \frac{\sinh ^2(  \widehat\rho / r)}{4 r^2}  \partial^\mu \theta \, \partial_\mu \theta   - {\widehat U}_{\rm out} ( \widehat\rho,\theta).
\end{align}
This difference in the potentials is crucial. They will take on the following forms depending on the domain, and thus lead to different phenomenology as we shall see shortly:
\begin{align}
 { \widehat U}_{\rm in} (  \widehat\rho ,\theta) &\equiv  {   U} \left(  r\tanh\frac{\widehat\rho}{r},\theta\right) ,
 \\
  { \widehat U}_{\rm out}  (  \widehat\rho ,\theta) &\equiv  { U} \left(r \coth\frac{\widehat\rho}{r},\theta\right) ,
 \end{align}
We also note that these potentials are similar to the ones for single-field $\alpha$-attractors, given in \eqref{Uinexact} and \eqref{Uoutexact}. 

Crucially, both \eqref{sfree1} and \eqref{sfree2} are singularity-free without any constraints on $\widehat \rho$ and $\theta$. As such, we have found that singular curves in multifield Lagrangians play a similar role to poles in single-field inflation: they effectively act as a barrier between pole-free models when dealing with Lagrangians which feature singularities, which reinforces their interpretation as ``model walls''. We therefore find that singularity-free models are once again the fundamental building blocks of generalised Lagrangians with singularities. 
We also note that the pole-free model \eqref{sfree1} corresponds to inflation with a  hyperbolic geometry, whereas the model~\eqref{sfree2} corresponds to inflaton on the punctured plane $\mathbb{R}^2/{(0,0)}$, since $\widehat\rho $ can never take on the value $\widehat \rho = 0$, since that would mean that it is on the singular curve. Topologically, this corresponds to a cylinder. The shape of the $U(\rho,\theta)$ potential will determine whether inflation can occur on either side of the model wall: if the non-canonical potential is convex at $\rho = r$, then the trajectory will roll away from the pole and inflation will occur in both regions (although we should expect that $r=\infty$ is reached in finite time in this case as well).

Before concluding, it is worth examining in brief the growth of  isocurvature perturbations on either side of the pole. 
Isocurvature (or entropic) perturbations (which are perpendicular to the trajectory) can feed into the usual curvature (or adiabatic) perturbations (which are parallel to the trajectory). This is known as \emph{entropy transfer}, and may cause the scalar amplitude $P_\mathcal{R}$ to grow even after crossing the horizon as $P_\mathcal{R}  = (1+T_\mathcal{RS}^2)P_\mathcal{R*} $. For two-field models, this transfer is usually encoded in the transfer functions $T_\mathcal{RS}$ and $T_\mathcal{SS}$, evaluated after horizon crossing. The former function tells us how much entropy transfer has occurred, whereas the latter tells us how much the entropy amplitude $P_\mathcal{S}$ grows, as $P_\mathcal{S}  = (1+T_\mathcal{SS}^2)P_\mathcal{S*} $. 

An important parameter for entropy transfer is the \emph{turn rate} $\omega$ (intuitively, as the trajectory turns in field space, perturbations that begin parallel to the trajectory obtain a perpendicular component). However, an important component in whether perturbations will grow is whether inflation occurs on a ridge or a valley. For two-field models, this is controlled by the slow-roll parameter~$\eta_{ss}$~\cite{Kaiser:2012ak}, which roughly corresponds to the second derivative perpendicular to the trajectory for flat field space. In this case, $\eta_{ss} < 0$ corresponds to a hilltop, in which case, entropy perturbations grow due to the tachyonic instability. Conversely, $\eta_{ss}>0$ corresponds to a valley, and $\eta_{ss}>1$ means that entropy perturbations will generally be dampened. However, for a curved field space, $\eta_{ss}$ picks up a term proportional to~$\epsilon S$, where~$S$ is the Ricci curvature of the field space \cite{Karamitsos:2017elm}. We may calculate this using the standard differential geometric expression with the help of the metric $G_{AB}$ and the Christoffel symbols:
\begin{align}
\begin{aligned}
S  &= - \frac{2}{r^2} \left[1- \text{csch}^2 \left(\frac{\hat \rho}{r} \right) \right].
\end{aligned}
\end{align}
We note that the curvature is positive for $|\widehat\rho/r|  \le  \text{csch}^{-1} \, 1 \approx 0.881$. Therefore, entropy growth will be suppressed well within in the region $|\widehat\rho/r| < 0.881$ but amplified far outside of it. Outside of the poles however, the curvature of the field space is constant (and negative). The amount to which entropy generation will be suppressed or amplified, however, will depend on the potential, which will be starkly different between the poles. Further study with specific forms for the potentials is required to calculate the degree to which isocurvature perturbations evolve depending on which domain the fields are found.

\section{Conclusions}
\label{conclusions}

Inflationary attractors are an interesting class of models in good agreement with observations. The formalism of pole inflation allows us to extract predictions from models which feature singular kinetic terms. However, the behaviour of models beyond the poles is not always given the attention it deserves, especially in light of the geometrical features that these singularities induce. Using the language of differential geometry, we have shown that models with poles correspond to manifolds which cannot be parametrised by a single chart. Indeed, a collection of charts (an atlas) is required, splitting the model into different domains. This has the physical implication that theories with poles can be viewed as a union of pole-free theories. We come to the conclusion that the domain on which the scalar field begins its evolution is equivalent to choosing a specific model.

We have shown that the interval in which the non-canonical field evolves can have a direct phenomenological impact. Specifically, we have demonstrated that if the potential is monotonic across the pole, then there is a switch between early- and late-time acceleration depending on the interval. We have shown this explicitly for monomial potentials, where inflation between the poles is in a sense dual to the late-time acceleration that occurs beyond the poles. The relation between the gradient of the potential and early- or late-time acceleration can also be seen in quintessential inflation, which features a monotonically decreasing potential. 
Thanks to this shape, $\alpha$-attractors featuring an exponential potential lead to a scenario in which the inflaton, after inflating and reheating, can reach the boundary of the field space in finite time. This scenario is geometrically non-trivial: it is exactly equivalent to inflation on the projective ray. Fundamentally, this indicates that the theory needs to be extended in order to capture the behaviour of the field at late times. However, it is possible that the Hubble drag prevents the field from reaching the boundary. In this case, we find that quintessence may be achieved without the need for introducing an offset to the potential. 

Multifield theories introduce a subtlety: certain singularities may only appear due to our parametrisation of the fields, much like coordinate singularities in general relativity. However, if we consider a coordinate-independent quantity (such as the line element), we can discriminate between coordinate and real singularities. To do so, we simply must determine whether there are any finite paths between points on either side of the singularity. 
The shortest of these between two points will by definition be a geodesic, and if such one exists and is on-shell as well (since the Hubble drag and the potential term will cause a deviation from the geodesic), then the singularity is only a coordinate one. We offer two examples: the Eisenhart lift, which features a coordinate singularity, and a two-field $\alpha$-attractor-like model with a real singularity. In the latter, we find two domains: one corresponding to a disc with hyperbolic geometry, and one corresponding to cylinder. By studying the effect of the curvature of these two domains on the growth of isocurvature perturbations, we qualitatively argue that they are amplified when outside of the hyperbolic disc. A more detailed study of the genuine multifield effects on either side of the singularity could shed some light as to whether it is possible to observationally distinguish in which domain we find ourselves.
  
The question of initial conditions is one that any model of inflation must contend with, and the phenomenological differences between the different domains (within and beyond the poles) raise quite a few questions regarding the initial conditions of attractor models. In most studies of inflation featuring a plateau, it is assumed that the inflaton starts up sufficiently high up the flat region in order to allow the $60$ or so $e$-foldings necessary to produce the observed features of the CMB today. This represents a kind of fine-tuning that is not always trivial to overcome \cite{Brandenberger:2016uzh}. It has been argued that the presence of plateaus in attractor models (after canonicalisation) can resolve the problem of initial conditions, as it is more likely that the field will start on the infinite plateau as opposed to the finite smaller region \cite{Linde:2017pwt}. Since it is not possible to  imposing a uniform probability density on an infinite interval \cite{Finn:2018krt}, it might make more sense to argue for a uniform distribution in the non-canonical field within the poles. However, in this case, we are tasked with explaining why the non-canonical field lives in that domain. It therefore looks as if choosing the initial value of the non-canonical fields represents a new kind of initial condition tuning. The choice of a domain now corresponds to different \emph{models} (as opposed to initial conditions \emph{within} a given model). Therefore, before we can consider the problem of initial conditions for a given \emph{Lagrangian}, it is very important to check whether said Lagrangian is hiding more than one model that may well have different predictions.
  
When we write down a Lagrangian (motivated either by phenomenological model building or particle theory), different initial conditions may lead to sufficient or insufficient inflation. However, if our Lagrangian features poles, we find that the choice of initial conditions also corresponds to the choice of a distinct canonical model. Therefore, any arguments that can be employed to resolve the problem of initial conditions in a pole-free canonical Lagrangian cannot also be used to argue why the original Lagrangian (if singular) reduces to said model. If we take a Lagrangian with poles as our starting point, then we must be prepared to argue why inflation must occur within a particular interval before we can attempt to solve the problem of initial conditions there. While no parametrisation is privileged over any other, the fact that different parts of field space are disconnected in the presence of poles is independent of any parametrisation we use (hence the need for an atlas). It seems that before attacking the problem of initial conditions, we must make a choice as to which  Lagrangian we consider more ``fundamental''. Investigating how the initial condition problem fares for Lagrangians that encode more than one model is an interesting question in its own right, and represents a potential avenue for further research.

\acknowledgments

The author would like to thank John McDonald, Kieran Finn, Apostolos Pilaftsis, Evangelos Sfakianakis,  S\'{e}bastien Renaux--Petel, Charlotte Owen and Konstantinos Dimopoulos for valuable comments and discussion. This work was supported by STFC via the Lancaster--Manchester--Sheffield Consortium for Fundamental Physics (ST/L000520/1).

\end{document}